%% file: manuscript.tex
\documentclass[conference]{IEEEtran}
\IEEEoverridecommandlockouts

\ifCLASSINFOpdf
\else
\fi

\usepackage[left=1.59cm,right=1.58cm,top=1.78cm]{geometry}
\usepackage{lipsum}
\usepackage{caption}
\usepackage{subcaption}
\usepackage{cite}
\usepackage{times}
\usepackage{graphicx}
\usepackage{algorithm}
\usepackage{algpseudocode}
\usepackage{amsmath}
\usepackage{stfloats}
\usepackage{caption}
\usepackage{color}
\usepackage{fancyvrb}
\usepackage{amsmath}
\usepackage{amssymb}
\usepackage{makeidx}
\usepackage{mdwmath}
\usepackage{mdwtab}
\usepackage{moreverb}
\usepackage{fancyvrb}
\usepackage{etoolbox}
\usepackage{mathtools}
\usepackage{mathptmx}
\usepackage{multirow}
\usepackage{color}
\usepackage{url}
\usepackage{tabularx}
\usepackage{enumitem}
\usepackage{hyperref}
\usepackage{booktabs}
\usepackage[T1]{fontenc}       
\usepackage[utf8]{inputenc}    
\usepackage{graphicx}          
\usepackage{amsmath,amssymb}   
\usepackage{tikz}              
\usetikzlibrary{positioning}   
\usetikzlibrary{calc}  
\usepackage{footnote}
\usepackage{rotating}
\usepackage{adjustbox}

\makesavenoteenv{figure}

\usepackage{xcolor}

\usepackage{titlesec}
\titlespacing\section{0pt}{12pt plus 4pt minus 2pt}{2pt plus 2pt minus 2pt}
\titlespacing\subsection{0pt}{12pt plus 4pt minus 2pt}{2pt plus 2pt minus 2pt}

\usepackage[nolist]{acronym}

\input{Acronyms}

\begin{document}



\title{AI-Powered Conflict Management in Open RAN: Detection, Classification, and Mitigation}


\author{\IEEEauthorblockN{Abdul Wadud\IEEEauthorrefmark{1}\IEEEauthorrefmark{2},~\IEEEmembership{Graduate Student Member,~IEEE}
Fatemeh Golpayegani\IEEEauthorrefmark{1},~\IEEEmembership{Senior Member,~IEEE} and} 
\IEEEauthorblockN{Nima Afraz \IEEEauthorrefmark{1},~\IEEEmembership{Senior Member,~IEEE}}
\IEEEauthorblockA{\IEEEauthorrefmark{1}School of Computer Science,
University College Dublin, Ireland}
\IEEEauthorblockA{\IEEEauthorrefmark{2}Bangladesh Institute of Governance and Management, Dhaka, Bangladesh}
\thanks{Corresponding author: Abdul Wadud (email: abdul.wadud@ucdconnect.ie).}}

%



\IEEEtitleabstractindextext{%
\begin{abstract}
Open \ac{RAN} was designed with native \ac{AI} as a core pillar, enabling \ac{AI}-driven \acp{xApp} and \acp{rApp} to dynamically optimize network performance. However, the independent \ac{ICP} adjustments made by these applications can inadvertently create conflicts-- direct, indirect, and implicit, which lead to network instability and \ac{KPI} degradation. Traditional rule-based conflict management becomes increasingly impractical as Open \ac{RAN} scales in terms of \acp{xApp}, associated \acp{ICP}, and relevant \acp{KPI}, struggling to handle the complexity of multi-\ac{xApp} interactions. This highlights the necessity for AI-driven solutions that can efficiently detect, classify, and mitigate conflicts in real-time. This paper proposes an \ac{AI}-powered framework for conflict detection, classification, and mitigation in Open \ac{RAN}. We introduce GenC, a synthetic conflict generation framework for large-scale labeled datasets with controlled parameter sharing and realistic class imbalance, enabling robust training and evaluation of \ac{AI} models. Our classification pipeline leverages \acp{GNN}, Bi-LSTM, and SMOTE-enhanced GNNs, with results demonstrating SMOTE-GNN's superior robustness in handling imbalanced data. Experimental validation using both synthetic datasets (5-50 xApps) and realistic ns3-oran simulations with OpenCellID-derived Dublin topology shows that AI-based methods achieve $3.2\times$ faster classification than rule-based approaches while maintaining near-perfect accuracy. Our framework successfully addresses \ac{ES}/\ac{MRO} conflict scenarios using realistic ns3-oran and scales efficiently to large-scale xApp environments. By embedding this workflow into Open \ac{RAN}'s \ac{AI}-driven architecture, our solution ensures autonomous and self-optimizing conflict management, paving the way for resilient, ultra-low-latency, and energy-efficient 6G networks.
\end{abstract}

\begin{IEEEkeywords}
AI, Open RAN, xApp Conflicts, Graph Neural Networks, Conflict Classification, Real-time Systems
\end{IEEEkeywords}}

\maketitle

\thispagestyle{plain}
\pagestyle{plain}

\IEEEdisplaynontitleabstractindextext

%
\IEEEpeerreviewmaketitle

\acresetall 

\section{Introduction}
\label{Sec:intro}
Open \ac{RAN} is increasingly recognized as a promising framework for next-generation communication systems, as it decouples software and hardware while enabling a broader and more flexible ecosystem. A key concept in Open \ac{RAN} is the native integration of \ac{AI}, where intelligent functionalities are embedded directly into the Open \ac{RAN} architecture rather than added as external modules in the form of \acp{xApp}, \acp{rApp}, and \acp{dApp}. In this vein, the O-RAN Alliance has recently released a comprehensive research report outlining a vision for \emph{Native AI in O-RAN} \cite{ORAN_NGRG_Research_Report_2023}. This document details how \ac{AI}-based \acp{xApp} can be employed to control and optimize network \acp{ICP}, particularly in the \ac{RIC}-thus unlocking new levels of automation and performance in 6G networks. However, in 6G’s \ac{AI}-native \ac{RAN}, where dozens of vendor \acp{xApp} will run concurrently over shared control parameters. Unmanaged conflicts in such environment can directly threaten reliability, automation, and SLA guarantees, making scalable conflict management a foundational requirement for 6G.

Despite these advances, the scope of the O-RAN Alliance report focuses primarily on leveraging \ac{AI} to improve network control and optimization with minimal attention to the inherently multi-dimensional challenges of conflict management, stating briefly about conflict mitigation in Section 6.2.3 of the technical specification in \cite{ric_oran_alliance}. In other words, while Open \ac{RAN}’s native \ac{AI} functionality may prevent \acp{xApp} from issuing conflicting commands in theory, there remains a critical gap in addressing how to detect, classify, and resolve these conflicts, especially when a large number of \acp{xApp} from various vendors simultaneously interact with thousands of \acp{ICP} and \acp{KPI}. This gap becomes even more pronounced as networks transition to 6G and the operational complexity of Open \ac{RAN} continues to grow. We dive deep into this matter in Section~\ref{sec:ProbForm}. 

Managing conflicts among different \acp{xApp} is crucial for ensuring smooth and efficient communication \cite{wadud2024qacm,adamczyk2023detection}. Therefore, this research focuses on \ac{xApp} conflict detection and classification in Open \acp{RAN}, where different software applications called \acp{xApp} control aspects like power distribution, resource allocation, and mobility management. However, as these \acp{xApp} operate on shared \acp{ICP}, their independent decisions introduce conflicts that degrade system \acp{KPI} \cite{wadud2024xapp, del2024pacifista, giannopoulos2025comix}. 

O-RAN Alliance's technical specification has classified \ac{xApp} conflicts into direct, indirect, and implicit categories \cite{ric_oran_alliance}. Previous studies often relied on manual detection or rule-based heuristics for conflict resolution \cite{adamczyk2023detection, adamczyk2023conflict, wadud2024xapp}. While these methods are effective for small-scale Open \ac{RAN} deployments, they become impractical in large-scale Open \ac{RAN} networks comprising a large number of \acp{xApp} and thousands of \acp{ICP}. In such scenarios, real-time conflict detection, classification, and mitigation must scale efficiently to meet strict latency constraints, which makes \ac{AI}-driven classification essential.

\begin{figure}[!ht]
 \centering
\includegraphics[scale=0.35]{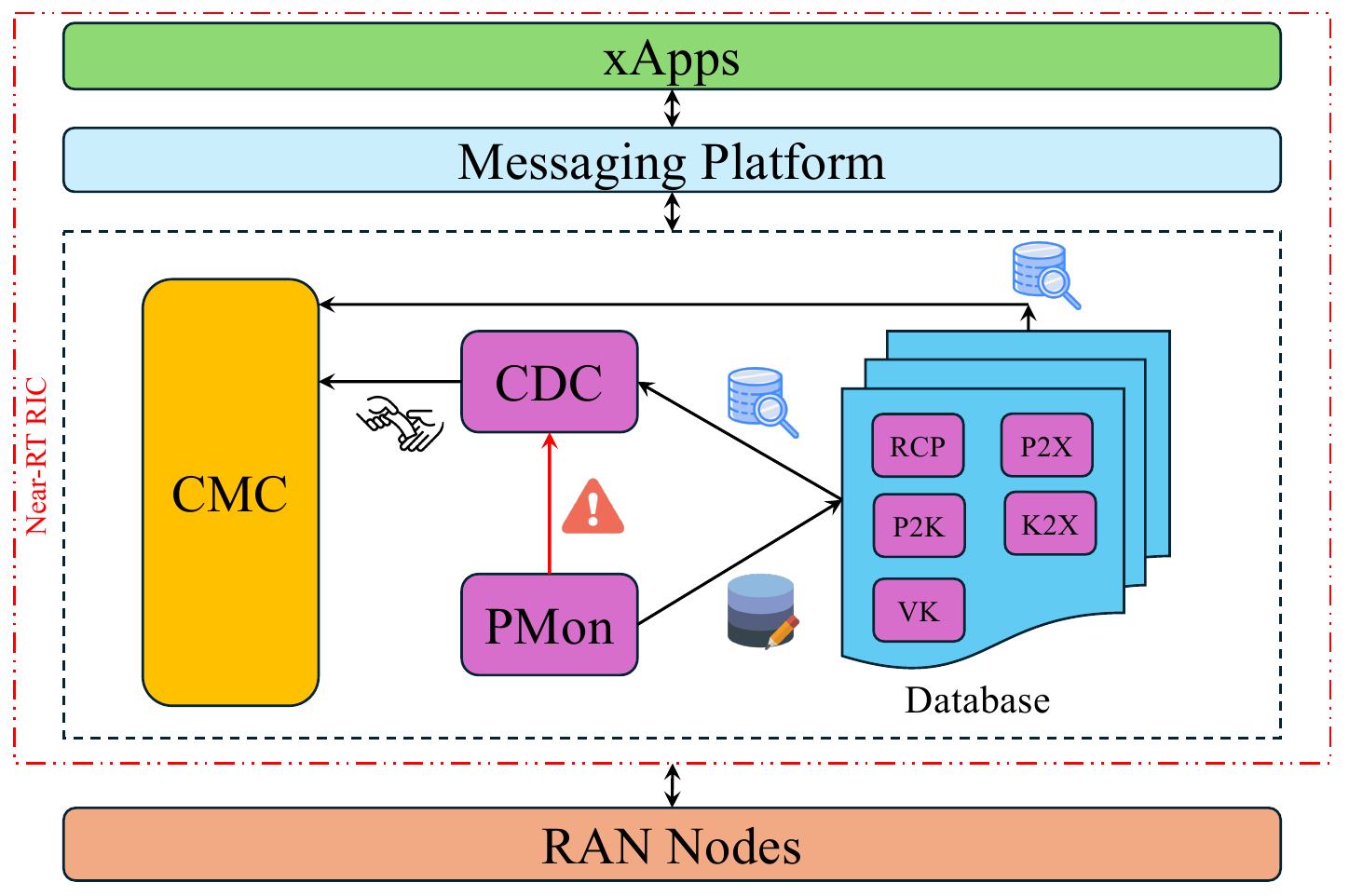}
	\caption{Conflict Management System in the \ac{Near-RT-RIC}.}
	\label{fig:cms_detection}
 \vspace{-0.1in}
\end{figure}

This work builds upon our previous works \cite{wadud2024qacm, wadud2023conflict}, which are focused on conflict mitigation strategies and introduced a \ac{CMS} framework, as shown in Fig.~\ref{fig:cms_detection}, for detection and mitigation of conflicts in the \ac{Near-RT-RIC}. Here, we emphasize the importance of accurate classification as a prerequisite for effective conflict resolution. The components of the \ac{CMS} are discussed briefly in Section~\ref{sec:cms} of this paper. Our results show that manual or rule-based classifiers deployed in the \ac{CDC} has higher classification latency in large Open \ac{RAN} networks due to the exponential complexity of \ac{xApp} interactions. Consequently, we propose an \ac{AI}-based classification framework that learns conflict patterns, enabling proactive and scalable conflict detection, classification and mitigation. Table~\ref{tab:traditional_vs_ai} compares the benefits of \ac{AI}-based approaches over traditional rule-based method based on state-of-the-art.

This paper advances conflict management in Open \ac{RAN} by addressing detection, classification, and mitigation of \ac{xApp} conflicts with scalable \ac{AI}-driven approaches. Our contributions are as follows:

\begin{itemize}
    
    \item We develop a stochastic simulation framework, GenC, to generate large-scale, labeled conflict datasets across various \ac{xApp} settings. Unlike real-world testbeds, which rarely capture all conflict types, GenC ensures controlled and reproducible conflict scenarios--crucial for benchmarking \ac{AI}-based solutions. It is an extension of the stochastic model used in previous studies \cite{wadud2023conflict, wadud2024qacm, zolghadr2024learning}.

    \item We implement and evaluate GNNs, Bi-LSTM, and SMOTE-enhanced GNNs for scalable conflict classification. We analyze their ability to handle imbalanced conflict datasets and demonstrate that SMOTE-GNN significantly enhances classification robustness compared to standard \ac{ML} models.

    \item We compare manual, rule-based conflict classification against \ac{AI}-driven models by highlighting the limitations of static heuristics in large-scale Open \ac{RAN} settings. We show that \ac{AI}-based methods outperform rule-based approaches in terms of scalability and computational efficiency, particularly under high \ac{xApp} density.

    \item We validate our AI-powered conflict management framework using realistic ns3-oran simulations with OpenCellID-derived Dublin topology, demonstrating $3.2\times$ speed improvement over rule-based methods while maintaining near perfect accuracy in practical \ac{ES}/\ac{MRO} \acp{xApp} conflict scenarios.

\end{itemize}

This work provides a comprehensive and future-proof framework for conflict management in Open \ac{RAN} by integrating synthetic conflict data, \ac{AI}-driven classification, and scalable mitigation strategies, paving the way for autonomous, \ac{AI}-native 6G networks.

\begin{table}[htbp]
\centering
\caption{List of Important Acronyms Used}
\begin{tabular}{ll}
\hline
\textbf{Acronym} & \textbf{Full Form} \\
\hline
CCO   & Capacity and Coverage Optimization \\
CDC   & Conflict Detection Controller \\
CDR   & Call Drop Rate \\
CIO   & Cell Individual Offset \\
CMC   & Conflict Mitigation Controller \\
DU    & Distributed Unit \\
DLT   & Downlink Throughput \\
EE    & Energy Efficiency \\
ES    & Energy Saving \\
ICP   & Input Control Parameters \\
K2X   & KPIs to Parameters Relationships \\
KPIs  & Key Performance Indicators \\
MLB   & Mobility Load Balancing \\
MRO   & Mobility Robustness Optimization \\
P2K   & Parameters to KPIs Relationships \\
P2X   & Parameters to xApp Relationships \\
PMon  & Performance Monitoring \\
RET   & Radio Electrical Tilt \\
RCP   & Recently Changed Parameter \\
RU    & Radio Unit \\
TL    & Traffic Load \\
TTT   & Time to Trigger \\
TxP   & Transmission Power \\
VK    & Violated KPIs \\
\hline
\end{tabular}
\label{tab:acronyms}
\end{table}


\begin{table*}[ht]
\centering
\caption{Comparison of Traditional Rule-Based vs. AI-Based Conflict Management in Open RAN}
\label{tab:traditional_vs_ai}
\renewcommand{\arraystretch}{1.2}
\begin{tabular}{|p{1.7cm}|p{5cm}|p{5.3cm}|p{5cm}|}
\hline
\textbf{Aspect} & \textbf{Traditional Rule-Based Methods} & \textbf{AI-Based Methods} & \textbf{Pros of Using AI-Based Approach} \\ \hline

\textbf{Conflict Detection} & 
\begin{minipage}[t]{\linewidth}
\raggedright
\begin{itemize}[leftmargin=*, noitemsep]
\item Reactive \ac{KPI}-based threshold monitoring \cite{adamczyk2023detection, wadud2024xapp, wadud2024qacm}
\item Detection occurs post-impact
\end{itemize}
\end{minipage}
&
\begin{minipage}[t]{\linewidth}
\raggedright
\begin{itemize}[leftmargin=*, noitemsep]
\item Proactive anomaly detection
\item Techniques: Autoencoders, Transformers, GNNs \cite{zolghadr2024learning}
\end{itemize}
\end{minipage}
&
\begin{minipage}[t]{\linewidth}
\raggedright
\begin{itemize}[leftmargin=*, noitemsep]
\item Identifies potential issues before major \ac{KPI} degradation
\end{itemize}
\end{minipage}
\\ \hline

\textbf{Conflict \ Classification} &
\begin{minipage}[t]{\linewidth}
\raggedright
\begin{itemize}[leftmargin=*, noitemsep]
\item Rule-based classification \cite{wadud2024xapp}
\item Not scalable with many \acp{xApp} and \acp{KPI}
\end{itemize}
\end{minipage}
&
\begin{minipage}[t]{\linewidth}
\raggedright
\begin{itemize}[leftmargin=*, noitemsep]
\item Scalable ML models (GNNs, LSTMs, etc.)
\item Learns dynamic relationships among \acp{xApp}
\end{itemize}
\end{minipage}
&
\begin{minipage}[t]{\linewidth}
\raggedright
\begin{itemize}[leftmargin=*, noitemsep]
\item Adapts  to new \acp{xApp}, \acp{ICP} and \acp{KPI} (non-disruptive model update)
\item Handles high-dimensional data without manual rules
\end{itemize}
\end{minipage}
\\ \hline

\textbf{Conflict     \ \ \  Mitigation} &
\begin{minipage}[t]{\linewidth}
\raggedright
\begin{itemize}[leftmargin=*, noitemsep]
\item Game-theoretic approaches (e.g., NSWF, EG) \cite{wadud2023conflict}
\item Complex at scale, less flexible for diverse \ac{QoS} \cite{wadud2024qacm}
\end{itemize}
\end{minipage}
&
\begin{minipage}[t]{\linewidth}
\raggedright
\begin{itemize}[leftmargin=*, noitemsep]
\item DRL, MARL \cite{zhang2022team}, Transformer-based optimizers
\item Real-time, SLA-aware resource allocation
\item online-RL (\ac{PPO}, 19\% throughput gain \cite{lee2021onlineRIC})
\end{itemize}
\end{minipage}
&
\begin{minipage}[t]{\linewidth}
\raggedright
\begin{itemize}[leftmargin=*, noitemsep]
\item Supports multi-objective optimization (\ac{QoS}, throughput, latency)
\item Learns from feedback to improve decisions over time
\item Adapts rapidly to changing traffic demands and network topologies
\end{itemize}
\end{minipage}
\\ \hline

\end{tabular}
\end{table*}

\section{Understanding Conflicts among xApps}
\label{sec:xApp_conflict}

The definitions of \ac{xApp} conflicts provided by O-RAN Alliance's technical specification \cite{ric_oran_alliance} are not comprehensive, not defined and explained clearly with proper example. Specifically, the subtle difference between indirect and implicit conflict is hard to understand. Therefore, we provide a concise but comprehensive definitions of these conflicts with real-\ac{xApp} examples in this section based on our recent work in \cite{wadud2024xapp}.

\begin{figure}[!ht]
 \centering
\includegraphics[scale=0.21]{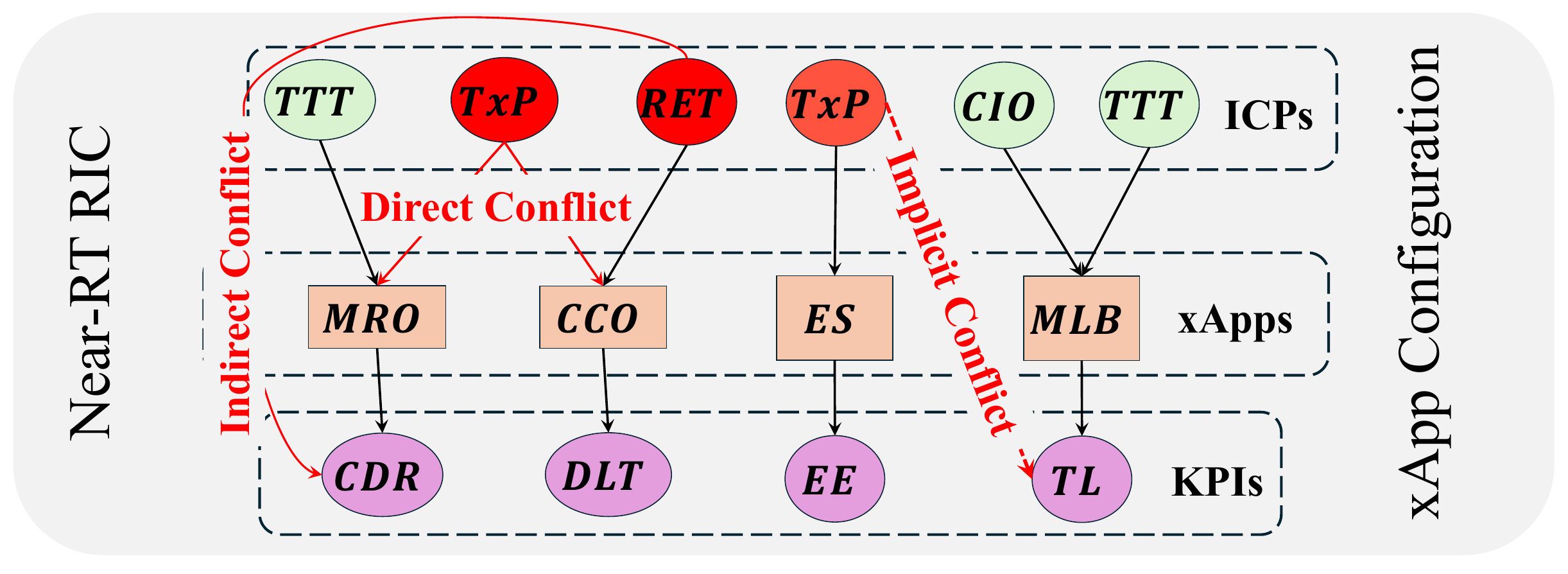}
	\caption{Various \ac{xApp} conflicts in Open RAN.}
	\label{fig:con_example}
 \vspace{-0.1in}
\end{figure}
\begin{itemize}
    \item \textbf{Direct Conflict:} This occurs when multiple \acp{xApp} explicitly control the same parameter that affects a network performance indicator. For example, if two applications try to adjust the same frequency band, their conflicting instructions may degrade signal quality. To explain this, we considered two \acp{xApp} \ac{MRO} and \ac{CCO} in Fig.~\ref{fig:con_example}. Both shares a common parameter \ac{TXP}, which may create a direct conflict between these two \acp{xApp} if conflicting setting for \ac{TXP} is instructed from them in subsequent request. 

    \item \textbf{Indirect Conflict:} This happens when an \ac{xApp} modifies a parameter that is not directly controlled by another \ac{xApp} but still affects a \ac{KPI} managed by that other \ac{xApp}. Imagine adjusting the transmission power of a base station, which in turn influences another \ac{xApp} that manages user connectivity. To explain this, we considered the \ac{MRO} \ac{xApp} in Fig.~\ref{fig:con_example}, where its \ac{KPI} \ac{CDR} is degraded when changes are made to the \ac{RET} parameter. Notably, \ac{RET} is not an \ac{ICP} for the \ac{MRO} \ac{xApp}; instead, it is an \ac{ICP} associated with the \ac{CCO} \ac{xApp}. This demonstrates how modifications by one \ac{xApp} can indirectly impact the performance of another. 

    \item \textbf{Implicit Conflict:} An implicit conflict arises when an \ac{xApp} changes a parameter that is not explicitly linked to the affected \ac{KPI} but still causes unexpected network behavior. This could be due to hidden dependencies between the changed parameter and the affected \ac{KPI} or something else within the system that are not immediately visible. To explain this, we considered the \ac{xApp} \ac{ES} and \ac{MLB} in Fig.~\ref{fig:con_example}, where \ac{MLB} detects high \ac{TL} on a particular cell and attempts to offload users to neighboring cells by lowering \ac{CIO}, making those neighboring cells more attractive. On the other hand, \ac{ES} autonomously reduces \ac{TXP} in the neighboring cells to conserve power, unaware that \ac{MLB} is actively trying to push users into those cells. As a result, many offloaded users experience poor signal strength and higher handover failures, leading to degraded \ac{QoS}, causing an implicit conflict. Here, \ac{TL} is the key metric affected by the implicit interaction between \ac{ES} and \ac{MLB}, illustrating how such conflicts can degrade network performance. 
\end{itemize}

A real-world simulation environment that captures all the aforementioned types of conflicts is extremely difficult to construct due to several practical limitations. First, Open RAN deployments require the integration of multiple independently developed \acp{xApp}, each designed to control specific aspects of the network such as mobility, power, or scheduling. Since most of these \acp{xApp} are not open source, developing them for large scale simulation are time consuming and not feasible. Even when multiple \acp{xApp} are deployed in the \ac{Near-RT-RIC}, the manifestation of actual conflicts--especially diverse types like indirect or implicit conflicts--is not guaranteed. Moreover, inducing all possible conflict types in a real system would require exhaustive and carefully timed parameter manipulations under varying network loads and topologies, which is neither scalable nor feasible in practice. Safety, reliability, and operational policies also limit the scope for experimentation in live networks, making it nearly impossible to observe edge-case conflicts or rare cascading failures.

To overcome these challenges, our previous work \cite{wadud2023conflict, wadud2024qacm} introduced a stochastic and data-driven simulation framework that \emph{mimics real-world scenarios} \cite{banerjee2021toward} and \emph{generates synthetic conflict data}. Initially designed with five \acp{xApp}, the framework has since been extended and generalized to support a larger number of \acp{xApp}, \acp{ICP}, and \acp{KPI}, enabling it to reproduce complex and varied conflict scenarios. This \emph{synthetic conflict generation (GenC)} method offers a scalable, controlled, and reproducible environment for studying conflict behaviors (See in Section~\ref{sec:genc}). It enables rigorous experimentation and benchmarking of \ac{AI}-driven conflict detection and resolution techniques, which would otherwise be unachievable with current real-world testbeds.

\section{Conflict Management in Open RAN}
\label{sec:cms}
The \ac{CMS} contains three main components for detection and mitigation of \ac{xApp} conflicts in the \ac{Near-RT-RIC}. These components are- the \ac{PMon}, the \ac{CDC}, and the \ac{CMC}. The \ac{PMon} component informs any \ac{KPI} anomalies to the \ac{CDC}, afterwards, it classifies the conflict through retrieving relationship information like- \ac{P2K}, \ac{P2X}, \ac{K2X}, and \ac{VK} from the \ac{Near-RT-RIC} database and passes the details to the \ac{CMC} for mitigation. Figure.~\ref{fig:cms_detection} illustrates \ac{CMS} architecture within the \ac{Near-RT-RIC}. The database components illustrated in this figure are mainly used for conflict detection purpose in addition to the mitigation related components discussed in our previous works \cite{wadud2023conflict, wadud2024qacm}. As this paper mainly focuses on detection and classification of \ac{xApp} conflicts, we only discuss detection related database components. The following provides discussion on each of the components one by one:

\textbf{PMon:} This component streams \ac{KPI} data from the \ac{Near-RT-RIC} database, and compares every KPI against its \ac{QoS} target. Any breach is logged in the \ac{VK} component of the database, and immediately triggers the \ac{CDC} to check whether \ac{xApp} actions are clashing. \ac{VK} was formerly known \ac{KDO} in our previous work \cite{wadud2023conflict}. 

\textbf{CDC:} When a \ac{KPI} breach alert arrives from \ac{PMon}, the \ac{CDC} reviews the latest parameter changed by any \ac{xApp}. By evaluating the \ac{KPI}-\ac{ICP}-{xApp} relationships using either rule-based (see Algortihm~\ref{alg:rule_based_mitigation}) or AI-driven detection methods, the \ac{CDC} determines the type of \ac{xApp} conflict occurred. Afterwards, every conflict type, together with the conflicting \acp{xApp} and \ac{ICP}, is logged and relayed to the \ac{CMC} for mitigation.

\textbf{CMC:} The \ac{CMC} treats every conflict either as a cooperative game \cite{wadud2023conflict, wadud2024qacm} or as priority \cite{adamczyk2023conflict} cases to mitigate the effect of these conflicts. Once the \ac{CDC} reports a conflicted parameter and violating \ac{KPI}, the \ac{CMC} gets triggered and starts its operation. 

\textbf{Database Components:} 
\begin{itemize}
    \item \textbf{RCP:} \ac{RCP} is a timestamped log inside the database that records every \ac{ICP} an \ac{xApp} has modified. It preserves a complete time-ordered history of recently changed parameters.
    
    \item \textbf{P2X:} \ac{P2X} represents the mapping between each \ac{ICP} and the set of \acp{xApp} that utilize or manage it. This relationship helps identify which applications might be affected when a parameter changes.
    
    \item \textbf{P2K:} \ac{P2K} defines the influence of a specific \ac{ICP} on one or more \acp{KPI}. It captures how adjustments in system parameters affect network performance metrics.
    
    \item \textbf{K2X:} \ac{K2X} indicates which \ac{xApp}s are responsible for maintaining or optimizing a given \ac{KPI}. It helps trace the source of performance degradation when a \ac{KPI} violation occurs.
    
    \item \textbf{VK:} \ac{VK} stores the list of KPIs that have breached their service-level thresholds following a parameter change. It is critical for identifying and classifying conflicts triggered by \ac{xApp} actions.
\end{itemize}

\begin{figure}[!ht]
 \centering
\includegraphics[scale=0.24]{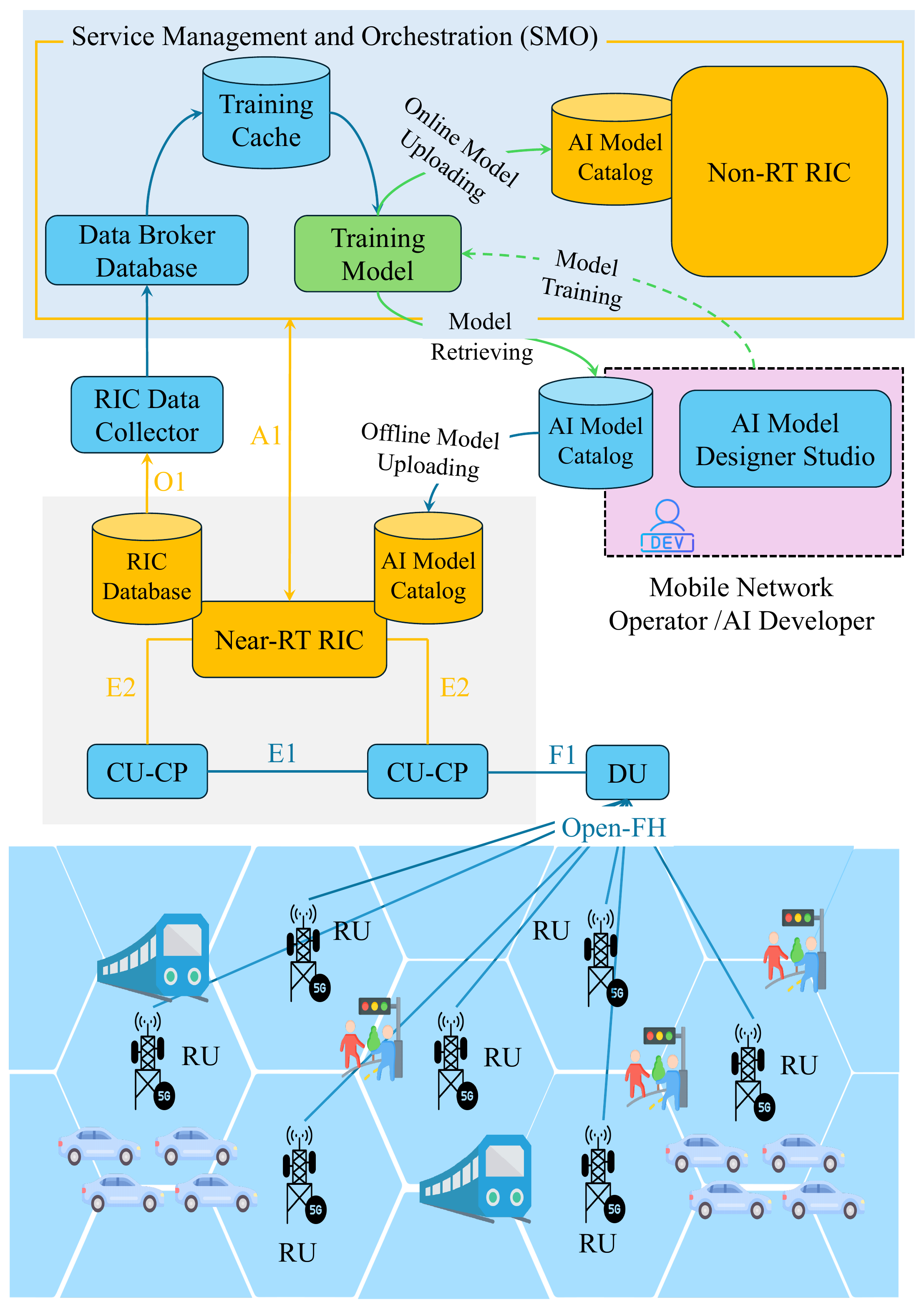}
	\caption{\ac{AI}-Powered Conflict Management Control-Loop in Open RAN}
	\label{fig:control_loop}
 \vspace{-0.1in}
\end{figure}

\begin{figure}[!ht]
 \centering
\includegraphics[scale=0.2]{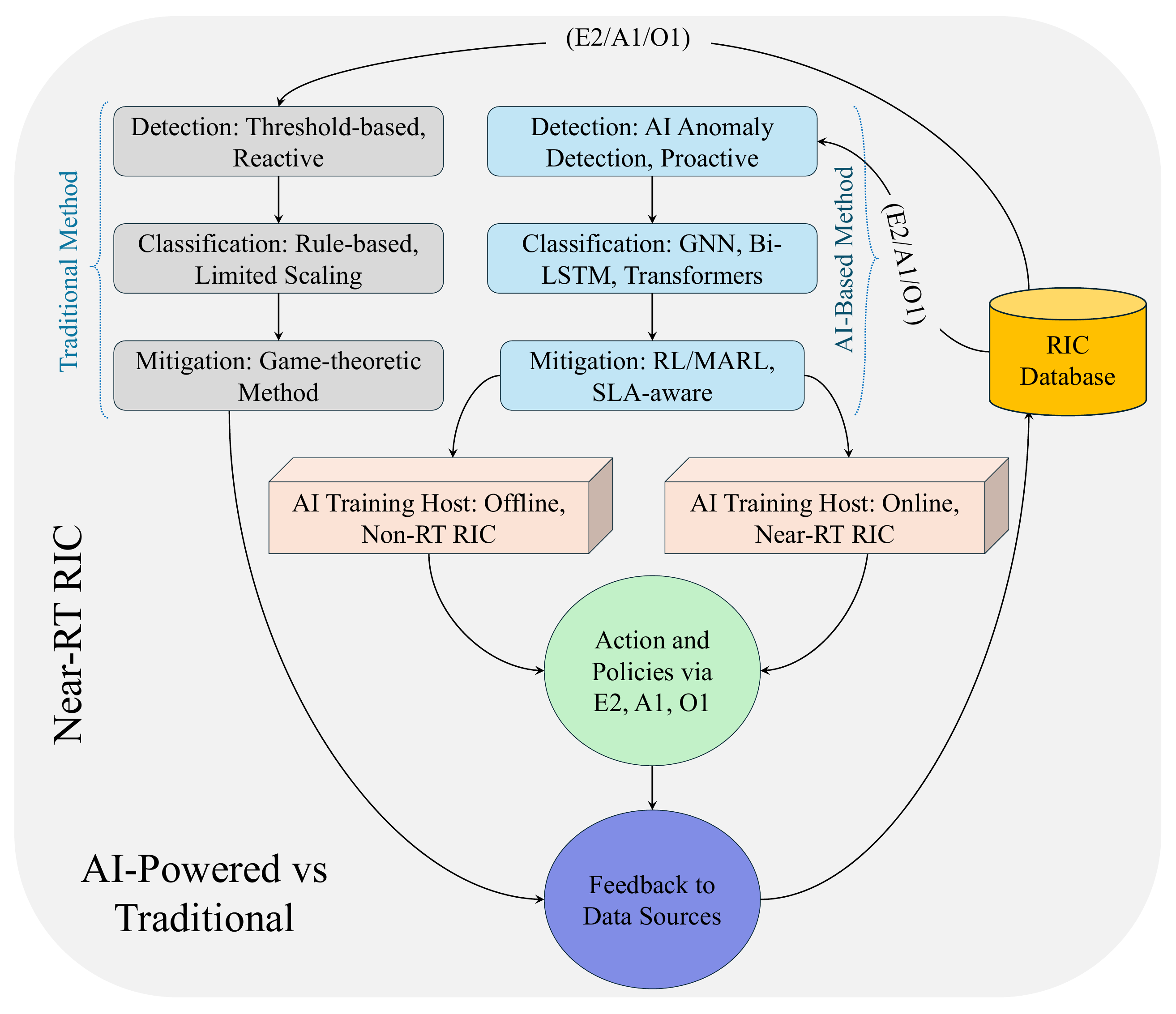}
	\caption{Traditional Rule-Based Versus \ac{AI}-Based Conflict Management Cycle in O-RAN}
	\label{fig:diff_control_loop}
 \vspace{-0.1in}
\end{figure}

\section{System Model and Problem Formulation}
\label{Sec:sysMpF}
\subsection{System Model}
\label{Sec:sysModel}
We consider an Open~\ac{RAN} environment comprising:
\begin{itemize}
    \item A set of $m$ xApps $\mathcal{X} = \{x_1, x_2, \dots, x_m\}$ deployed on the RIC, each potentially from different vendors;
    \item A set of $n$ controllable \acp{ICP} $\mathcal{P} = \{p_1, p_2, \dots, p_n\}$, where typically $n \gg m$;
    \item A set of $k$ \acp{KPI} $\mathcal{K} = \{k_1, k_2, \dots, k_k\}$ reflecting network objectives and \ac{SLA}s.
\end{itemize}
Each xApp $x_i \in \mathcal{X}$ manipulates a subset of parameters $\mathcal{P}_i \subseteq \mathcal{P}$ and targets a subset of KPIs $\mathcal{K}_i \subseteq \mathcal{K}$. The global control state at time $t$ can be represented as a vector $\mathbf{p}(t) = [p_1(t), p_2(t), \dots, p_n(t)]$. The system's observed performance is measured by $\mathbf{k}(t) = [k_1(t), k_2(t), \dots, k_k(t)]$.

Let $A_{i,j} = 1$ if $x_i$ and $x_j$ act on overlapping parameter(s), i.e., $\mathcal{P}_i \cap \mathcal{P}_j \neq \varnothing$, indicating potential for conflict. The system dynamics can be expressed as:
\[
\mathbf{k}(t+1) = f\big(\mathbf{p}(t), \mathbf{w}(t)\big),
\]
where $f(\cdot)$ models the RAN evolution and $\mathbf{w}(t)$ represents exogenous factors (e.g., traffic, channel variation).

The control objective is to coordinate all $x_i \in \mathcal{X}$ such that the aggregate KPIs meet or exceed their respective SLAs:
\[
k_\ell(t) \geq \mathrm{SLA}_\ell, \quad \forall k_\ell \in \mathcal{K}.
\]
This abstract model captures the high-dimensional, multi-agent, and multi-objective nature of Open~RAN control.

\subsection{Problem Formulation}
\label{sec:ProbForm}
In the context of Open \ac{RAN} systems, the optimization problem grows significantly more complex as the number of \acp{xApp}, \acp{ICP}, and \acp{KPI} increases. Specifically, when a large number of \acp{xApp} from various vendors simultaneously interact with thousands of \acp{ICP} and \acp{KPI}, the solution space expands exponentially due to the combinatorial explosion of possible parameter configurations. Each \ac{xApp} introduces its own set of \acp{ICP} and optimizes for specific \acp{KPI} (e.g., throughput, latency, energy efficiency), which may conflict with the objectives of other \acp{xApp}. This creates a high-dimensional optimization problem characterized by intricate inter-dependencies and potential conflicts among \acp{xApp}, which requires sophisticated coordination to achieve \ac{SLA} goals. Furthermore, the computational complexity of evaluating and optimizing thousands of \acp{KPI} across heterogeneous \acp{xApp} poses significant scalability challenges, necessitating advanced algorithms to navigate the vast solution space efficiently.

To dive deep into the technicalities of the scaling, we pinpoint three specific reasons:
\begin{itemize}
    \item \textbf{Combinatorial interaction space:} A modern \ac{gNB} already exposes hundreds of tunable configuration leaves in each 3GPP-NR object (e.g., gnbDuFunction, cellCarrier, beamformingGroup), and the O-RAN YANG tree augments these with dozens more per leaf class; across all RAN functions, this easily exceeds $2000$ ICPs per cell \cite{ORANSC2020, ORANSC2025}. If $m$ xApps each touch even $n << 2000$ overlapping parameters, the potential pair-wise collision space grows as $O(m^2 \times n^2)$. For only $20$ xApps tweaking 5 shared parameters, the conflict matrix already has $\approx 10000$ entries to check every control interval. 
    
    \item \textbf{High-dimensional cause/effect ambiguity:} \acp{KPI} are rarely mapped one-to-one to a single parameter. With hundreds of correlated counters (e.g., BLER, CQI variance, PRB utilisation), a small adjustment in power-control might surface seconds later as mobility failures or slice \ac{SLA} violations. Classifying which parameter triggered the drift, therefore, becomes an under-determined inverse-problem in a \ac{KPI} space, something that adds huge computational overhead to a simple rule-based “conflict detection” method to resolve.
    
    \item \textbf{Multi-timescale instability:} \ac{Near-RT-RIC} \acp{xApp} loop every 10ms to 1s, while non-RT \acp{rApp} may reconfigure hourly. The larger the parameter set, the more likely two loops will touch the same parameter at different cadences. This is a classic control-theory “integrator wind-up”. Recent sandbox studies show that even two benign xApps (throughput-maximisation vs. energy-saving) can drive throughput down by 50\% once their decisions start oscillating \cite{del2024pacifista}. 
\end{itemize}

By addressing these challenges, our proposed AI-driven approach enables effective management of large-scale, multi-vendor xApp interactions, thereby enhancing system performance and scalability.

\section{AI-Powered Conflict-Management Control Loop}
\label{sec:aiControlLoop}

In this section, we discuss each component of our proposed AI-Powered Conflict-Management Control Loop. Figure~\ref{fig:control_loop} expands the high-level \ac{AIMLFW} blocks of the O\textendash RAN Alliance into a concrete, five-stage loop that detects, classifies and resolves \ac{xApp} conflicts while continuously learning from live data (see Sec.~\ref{subsec:online_learning} for how runtime model updates are performed). Table~\ref{tab:loopRoles} summarizes the key roles and artifacts.

\begin{table}[t]
  \caption{Roles and artifacts in the AI control loop}
  \label{tab:loopRoles}
  \centering
  \begin{tabularx}{\linewidth}{@{}>{\raggedright\arraybackslash}p{0.30\linewidth}X@{}}
    \toprule
    \textbf{Block} & \textbf{Main function in conflict management} \\
    \midrule
    RIC Data Collector & Streams \acp{KPI}, \acp{RCP}\,/\,logs from DU/CU to the training cache \\ \midrule
    Data Broker DB     & Stores time-aligned telemetry and contextual \ac{EI} \\ \midrule
    Non-RT-RIC (Training Cache \& Model Training) & Online training + periodic retraining of detection, classification and mitigation models Online updates follow the workflow in Sec.~\ref{subsec:online_learning}. \\ \midrule
    AI Model Designer Studio   & Developer option for designing and training online models and deploying their offline versions to the \ac{Near-RT-RIC} \\ \midrule
    AI Model Catalog   & Version control and governance for rApp/xApp models \\ \midrule
    Near-RT-RIC (Inference) & Runs trained models, issues E2 SET/INSERT commands \\ \midrule
    O-DU/O-RU          & Enforces parameter updates; sends fresh KPIs back to loop \\
    \bottomrule
  \end{tabularx}
\end{table}

\subsection{Step 1 – Data Collection and Contextualization}
The \textit{RIC Data Collector} subscribes to Open \ac{RAN} \ac{KPM} and \ac{RAN} Slice SM streams over the E2 interface and augments every record with contextual \ac{EI} such as the RCP, P2X and K2X look-ups and the VK list. All tuples are time-stamped and stored in the \textit{Data Broker Database}, providing a single source of truth for model training and forensics.

\subsection{Step 2 – On/Offline Model Training (Non-RT RIC)}
Batches of curated data are moved into the \textit{Training Cache}. Here an operator or a third-party AI developer invokes the \textit{AI Model Designer Studio} to (re)train three families of models:

\begin{enumerate}[label=(\alph*),leftmargin=*]
  \item \emph{Early-warning anomaly detector} (Sec.~\ref{sec:react2proact}),
  \item \emph{Conflict classifier} (Sec.~\ref{sec:aiClassification}),
  \item \emph{Policy optimiser / mitigator} (Sec.~\ref{sec:conflict_native_ai}).
\end{enumerate}

Each artifact is exported together with a JSON schema describing its
input features and latency budget, then placed in the \textit{AI Model
Catalog} for life-cycle management. A reference implementation that couples Kubeflow pipelines with KFServing to realize MLOps-level-1 retraining inside the \ac{AIMLFW} is reported in \cite{lee2021onlineRIC}. This pipeline also supports continuous learning. When sufficient volume of new samples becomes available, the Non-RT RIC triggers the retraining-and-redeployment process described in Sec.~\ref{subsec:online_learning}.

\subsection{Step 3 – Model On-Boarding and Governance}
The catalog exposes standard \ac{AIMLFW} \acp{API} for offline model
uploading.  When an updated inference bundle passes governance checks
(license, explainability, test-cover), it is signed and pushed to the
\ac{Near-RT-RIC}.  Version pinning ensures \acp{xApp} can roll back to a
known-good policy if unforeseen events occur.

\subsection{Step 4 – Near-RT Inference and Enforcement}
Inside the \ac{Near-RT-RIC}, the inference host executes the detector and classifier within the 10ms to 1s near-RT budget.  Upon a positive conflict verdict it calls the policy optimizer, which returns a new parameter vector.  An “enforcer” xApp converts that vector into an \texttt{E2-SET} (or \texttt{E2-INSERT}) message targeting the affected \ac{O-RU}/\ac{O-DU}.

\subsection{Step 5 – Closed-Loop Feedback and Continuous Learning}
Updated KPIs flow back to the \textit{RIC Data Collector}, allowing:

\begin{itemize}[leftmargin=*]
  \item performance verification of the mitigation;
  \item drift detection for the anomaly model;
  \item automated triggers for the next retraining cycle.
\end{itemize}
These KPI feedback streams form the basis for triggering online updates of the AI models as detailed in Sec.~\ref{subsec:online_learning}. If the measured KPIs still breach the SLA, the loop reiterates with an
alternative policy--or escalates to the SMO for a slower, O1-level
reconfiguration--thereby realizing an end-to-end, self-healing workflow.
This dedicated section keeps the architectural picture in one place and
provides the reader with a clear narrative before Section~\ref{sec:react2proact} zooms into each ML block.

\section{Traditional Rule-Based Versus AI-Based Conflict Management in Open RAN}

Conflict-management techniques can be viewed along \emph{two orthogonal dimensions}. \emph{Timing} distinguishes \textit{reactive} methods, which act only after a performance deviation is observed, from \textit{pro-active} methods, which explore or predict future states (e.g., via a digital twin or time-series forecasting) and act \emph{before} degradation occurs. \emph{Decision engine} distinguishes \textit{rule-based} controllers (static if-then logic, constraint solvers, formal verification) from \textit{AI-driven} controllers (statistical learning, reinforcement learning, graph inference). The resulting \(2\times2\) design space is summarized in Table \ref{tab:design_space}; all four quadrants are active research areas.

\begin{table}[t]
  \caption{Orthogonal design space for xApp conflict management}
  \label{tab:design_space}
  \centering
  \begin{tabularx}{\linewidth}{@{}lXX@{}}
    \toprule
                      & \textbf{Reactive (post-fault)} & \textbf{Pro-active (pre-fault)} \\ \midrule
    \textbf{Rule-based} &
      KPI-threshold alarms (Open \ac{RAN} \ac{PMon}) &
      Digital-twin “what-if” tests or formal model-checking before issuing \texttt{E2\,SET} commands \cite{i14y} \\[2pt]
    \textbf{AI-based} &
      Online-learning root-cause engines that update only after a KPI breach \cite{onlineLearningRCA} &
      Forecast/simulation models that predict conflict likelihood minutes ahead and optimize actions accordingly \cite{del2024pacifista} \\ \bottomrule
  \end{tabularx}
\end{table}
Our proposal occupies the \emph{AI + pro-active} quadrant, because large-scale, multi-vendor O-RAN deployments require both early warning and the ability to generalize beyond hand-crafted rules. Accordingly, Subsections~\ref{sec:react2proact} and \ref{sec:rule2adaptive} now refer to “rule-based reactive” and “AI-based pro-active” approaches only as \emph{illustrative instances} inside this design space--not as mutually exclusive categories. Having introduced the end-to-end \ac{AI}-powered control loop in Section~\ref{sec:aiControlLoop}, we now compare the core building blocks- detection, classification, and mitigation under both conventional rule-based and modern \ac{AI}-driven designs as illustrated briefly in Fig.~\ref{fig:diff_control_loop}. These approaches map directly to the inference modules highlighted in Fig.~\ref{fig:control_loop}.

\subsection{Conflict Detection: From Reactive Thresholding to Proactive Prediction}
\label{sec:react2proact}
\subsubsection{Conventional Approach: KPI Thresholding}
In many Open \ac{RAN} implementations, conflicts are detected by monitoring \ac{KPI} values such as latency, \ac{DL}/\ac{UP} throughput, or power consumption. A conflict triggers an alarm only when one or more \acp{KPI} violate predefined thresholds. Although straightforward, this approach is inherently reactive: by the time a threshold is violated, the conflict may already have inflicted significant performance degradation resulting in slower response times, increased call drops, and suboptimal user experiences.

\subsubsection{AI-Based Approach: Early Anomaly Detection}
\ac{AI} techniques can predict emerging conflicts by continuously analyzing historical and real-time data in search of subtle anomalies. Models like Autoencoders or Transformers learn the normal \textit{signature} of network behavior and flag any deviations as potential conflicts. By pinpointing these early warning signs, vendors/operators can intervene \emph{before} \ac{KPI} degradation becomes critical. This proactive detection will minimize performance impact and optimizes resource efficiency and operational costs. \ac{AI} models can be retrained periodically (runtime retraining workflow described in Sec.~\ref{subsec:online_learning}), ensuring they remain accurate as network conditions, traffic patterns, and the range of \acp{xApp} evolve. For example, in a vehicular network scenario, handover failures often serve as early indicators of potential conflicts. If an \ac{AI} model detects an increased handover failure rate for high-speed vehicles near cell edges--despite stable signal strength--this could indicate an \ac{xApp}-induced anomaly. An \ac{AI}-based predictor can analyze past mobility patterns and preemptively adjust handover parameters before large-scale \ac{KPI} degradation occurs, ensuring smoother transitions between cells and improved \ac{QoS}.


\begin{figure*}[!ht]
 \centering
\includegraphics[scale=0.6]{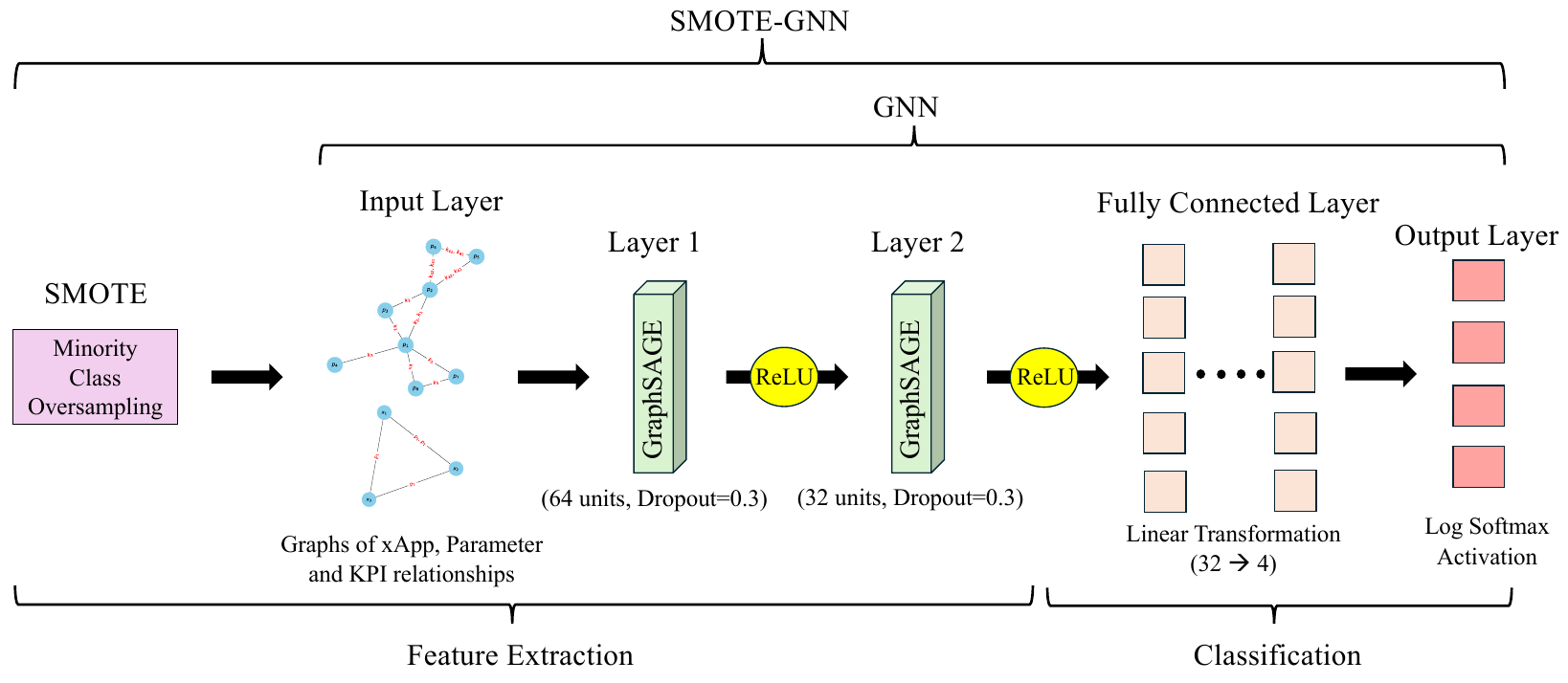}
	\caption{Architecture of GNN Method with and without SMOTE.}
	\label{fig:gnn_architecture}
 \vspace{-0.1in}
\end{figure*}

\subsection{Conflict Classification: From Rule-Based Logic to Adaptive Intelligence}
\label{sec:rule2adaptive}
\subsubsection{Traditional Approach: Static Rule-Based Systems}
Traditionally, conflict classification has been done by manually crafted rules. In this study, we develop the rule-based method illustrated in Algorithm~\ref{alg:rule_based_mitigation} to annotate various conflict types. While the method is easily interpretable, this approach becomes impractical when dealing with hundreds of \acp{xApp} from multiple vendors, each with its own \acp{KPI} and \acp{ICP}. Frequent updates to rules (to account for new \acp{xApp} and new relationships between \acp{KPI} and \acp{ICP}, along with the sheer number of possible conflict scenarios, make purely rule-based classification systems challenging to maintain at scale. The following discusses the Algorithm~\ref{alg:rule_based_mitigation}, its complexity, and why AI-based methods can be useful to be an alternative to this method. 


\begin{algorithm}
\caption{Rule-Based Conflict Annotation in Open RAN}
\label{alg:rule_based_mitigation}
\begin{algorithmic}[1]
\Require Retrieved dataset $\mathcal{D}$ from \ac{Near-RT-RIC} database at timestamp $t_{stamp}$ with \ac{xApp} interactions and network \acp{KPI}
\Require Mapping tables: $\mathcal{P}_{KPI}$ (P2K), $\mathcal{X}_{KPI}$ (K2X), $\mathcal{X}_{P}$ (P2X)
\Require Set of unassigned parameters $\mathcal{U}$

\State Load dataset $\mathcal{D}$ and initialize conflict counters
\For{each row $r \in \mathcal{D}$}
    \State Extract violating KPIs $V \leftarrow r[\text{VK}]$
    \State Extract instructing \ac{xApp} $X_i$ and changed parameter $P_c$ from $r[\text{RCP}]$
    \If{$V$ is empty \textbf{or} $X_i$ is empty \textbf{or} $P_c$ is empty}
        \State Label as \textbf{No Conflict} and continue
    \EndIf

    \For{each $kpi \in V$}
        \State Retrieve $P_k \leftarrow \mathcal{P}_{KPI}[kpi]$ \Comment{Affecting KPI}
        \State Retrieve $X_k \leftarrow \mathcal{X}_{KPI}[kpi]$ \Comment{xApps influencing KPI}
        \State Retrieve $X_p \leftarrow \mathcal{X}_{P}[P_c]$ \Comment{xApps controlling $P_c$}

        \If{$X_i \in X_k$ \textbf{and} $|X_k| = 1$}
            \State Label as \textbf{No Conflict} and continue
        \EndIf

        \If{$P_c \in P_k$}
            \If{$X_p \cap X_k \neq \emptyset$} 
                \State Label as \textbf{Direct Conflict}
            \Else
                \State Label as \textbf{Indirect Conflict}
            \EndIf
        \Else
            \If{$P_c \in \mathcal{U}$}
                \State Label as \textbf{Implicit Conflict}
            \Else
                \State Label as \textbf{No Conflict}
            \EndIf
        \EndIf
    \EndFor          
\EndFor              
\State Compute conflict statistics and mitigation strategies
\State Generate reports on conflict frequencies and computation times
\end{algorithmic}
\end{algorithm}

\noindent\textbf{Discussion of Algorithm~\ref{alg:rule_based_mitigation}:}  
Line\,1 \emph{loads} the retrieved log table $\mathcal D$ from the \ac{Near-RT-RIC} database and initializes counters. Line\,2 starts the outer loop that scans each of the $n$ rows.  Line\,3 parses three fields from the current row: the violated-KPI list $V$, the instructing xApp $X_i$, and the recently-changed parameter $P_c$. Lines\,4-6 implement an early exit: if any field is missing the row is labeled \textit{No-Conflict} and the algorithm proceeds to the next entry. Line\,7 launches an inner loop over every KPI in $V$ (average cardinality $\bar v$). Lines\,8-10 perform three constant-time hash look-ups: (i) $P_k$, the parameter group affecting the KPI, (ii) $X_k$, the set of xApps that manage that KPI, and (iii) $X_p$, the set of xApps that control $P_c$. Lines\,11-13 test whether $X_i$ is the \emph{only} manager of the KPI; if so the event is harmless and the algorithm skips to the next KPI. Line\,14 checks whether the changed parameter belongs to $P_k$. If it does, Lines\,15-17 decide between a \textit{Direct} versus \textit{Indirect} conflict by evaluating the set intersection $X_p\cap X_k$ (cost $O(|X_p|+|X_k|)$, denoted $\bar a$ on average). If $P_c\notin P_k$, Lines\,18-22 distinguish an \textit{Implicit} conflict (when $P_c$ is still unassigned) from \textit{No-Conflict}. The inner loop then advances to the next KPI and, after completion, Line\,23 returns to the next row of $\mathcal D$.

\medskip
\noindent\textbf{Overall cost:}  With $n=|\mathcal D|$, $\bar v=E[|V|]$ and $\bar a=E[|X_p|+|X_k|]$ the running time is
\[T_{\text{rule}} \;=\; O(n\bar v\bar a),
\qquad
S_{\text{rule}} = O(n)+|\text{P2K}|+|\text{K2X}|+|\text{P2X}|.\]
Here, $T_{\text{rule}}$ and $S_{\text{rule}}$ represent time and space complexities, respectively. This rule-based method, even for a modest configuration of 20 xApps ($\bar a\!\approx\!20$) and $\bar v=5$, if $\mathcal D$ contains one million log rows, then it already entails to $\sim$100 M set operations. This large number of operations can push inference latency well beyond the 10ms-1s near-RT budget. The space complexity $S_{\text{rule}} = O(n)+|\text{P2K}|+|\text{K2X}|+|\text{P2X}|$ assumes efficient hash table implementations for the mapping lookups, which is realistic for production RIC deployments.

\noindent\textbf{Why AI scales better:}  A \ac{GNN} or Bi-LSTM replaces the nested per-row branching with a single batched tensor operation whose per-row cost is $O(1)$ and maps efficiently to \ac{SIMD}/\ac{GPU} hardware.  Adding a new \ac{KPI} or \ac{xApp} only requires \emph{offline} model retraining, no rule rewrite, so the AI pipeline offers both lower latency and lower maintenance overhead when the system scales to hundreds of \acp{xApp} and thousands of \acp{ICP}.

\subsubsection{AI-Based Approach: Scalable ML Classification}
\label{sec:aiClassification}
\begin{figure*}[!ht]
 \centering
\includegraphics[scale=0.6]{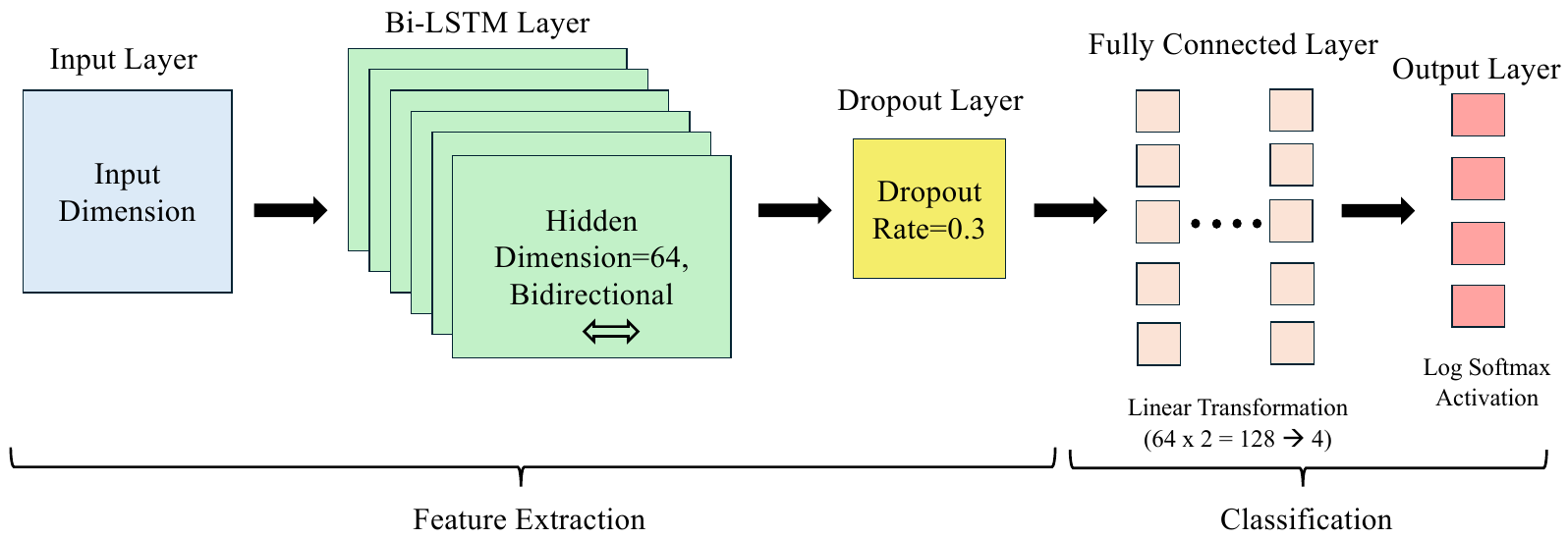}
	\caption{Architecture of Bi-LSTM Method.}
	\label{fig:lstm_architecture}
 \vspace{-0.1in}
\end{figure*}

\ac{AI}-driven classification solutions could tap into powerful models such as GNNs and Bi-LSTM networks, which excel at finding patterns across large, interlinked data sets. In an Open \ac{RAN} context, a GNN treat \acp{xApp}, Parameters, \acp{KPI}, Shared Parameters, and \ac{KPI}-Parameter groups as nodes in a graph, uncovering how one \ac{xApp}’s actions ripple through the network (see in Fig.~\ref{fig:gnn_architecture}). Bi-LSTM-based methods, on the other hand, can detect the temporal and sequential nature of conflicts by tracking changes in \ac{KPI} trends over time. Both techniques are far more adaptive and scalable than static rule-based systems, allowing the network to accommodate new \acp{xApp} and vendor solutions with minimal manual intervention. Architecture of GNN with or without SMOTE and Bi-LSTM is illustrated in Fig.~\ref{fig:gnn_architecture} and Fig.~\ref{fig:lstm_architecture}, respectively. 

\subsection{Conflict Mitigation: From Game Theory to AI-Driven Optimization}

\subsubsection{Conventional Approach: Game-Theoretic Solutions}
Once a conflict is detected and classified, operators have traditionally turned to game-theoretic frameworks like Nash’s Social Welfare Function (NSWF) or the Eisenberg-Gale (EG) method to allocate resources among conflicting \acp{xApp}. These techniques aim to balance fairness and efficiency but grow complex and computationally heavy as more \acp{xApp} come online. Ensuring each \ac{xApp}’s \ac{QoS} requirements are met further complicates the problem, especially if the network must respond in near real-time \cite{wadud2023conflict, wadud2024qacm}.

\subsubsection{AI-Based Approach: Intelligent Multi-Objective Optimization}
\ac{AI} algorithms, particularly those based on \ac{RL}, can resolve conflicts dynamically by learning from past decisions and feedback signals in real-world conditions. For instance, a Deep Q-Network (DQN) or Proximal Policy Optimization (PPO) agent can quickly propose resource allocation strategies that satisfy \ac{QoS} constraints while preserving overall network performance \cite{zhang2022team}. Transformers and attention-based models add another layer of flexibility by capturing complex inter-dependencies across \acp{xApp}. Unlike purely game-theoretic methods, \ac{AI}-driven solutions adjust in real time, scaling faster as the number of \acp{xApp} and traffic demands increase. 


\subsection{Relation to O-RAN WG2 and WG3 Conflict-Mitigation Efforts}
\label{sec:oran_wg_relation}

Recent O-RAN Alliance activities have begun formalizing conflict-management principles across different layers of the architecture. WG2 focuses on \emph{policy- and configuration-level} conflicts that arise before actions reach the Near-RT RIC. The WG2 Study on A1 Policy Conflict Mitigation~\cite{oran_wg2_plc_conflict_tr} specifies how overlapping or incompatible rApp policies can be detected and arbitrated using priorities or pre-emption. A complementary WG2 effort, the R1 Service-related Conflict Mitigation report~\cite{oran_wg2_r1servconfmit_tr}, describes pre-action conflict avoidance for O1/R1 configuration changes, including operator-defined arbitration rules (e.g., KPI priority or KPI targets) and avoidance strategies (accept one request, reject all, defer, or notify). These efforts focus on identifying \emph{potential} conflicts before actions are applied to the network.

WG3, in contrast, addresses \emph{runtime} conflicts that occur when multiple xApps concurrently issue E2-related control actions. The WG3 Conflict Mitigation technical reports~\cite{oran_wg3_confmit_tr,oran_wg3_confmit_draft} formalize direct, indirect, and implicit conflict types and define high-level procedures for conflict detection, resolution, and avoidance. These reports specify architectural behaviour and “solution requirements” but intentionally do not prescribe concrete machine-learning algorithms or provide evaluation methodologies.

Our CMS complements both WG2 and WG3 by providing the missing algorithmic layer for scalable runtime conflict handling. Using GenC-generated datasets and ns3-oran traces, we implement and evaluate GNN, Bi-LSTM-, and SMOTE-GNN-based classifiers capable of identifying conflicts at KPI/ICP level when dozens of xApps operate simultaneously. While WG2 and WG3 define \emph{what} conflicts must be handled within the O-RAN architecture, our CMS demonstrates \emph{how} such conflicts can be efficiently detected and classified in practice. Thus, the methods developed here can serve as candidate algorithmic realizations for the conflict-management capabilities anticipated, but not yet specified, in the current WG2 and WG3 outputs.

\section{Integrating Conflict detection, Classification, and mitigation into Open RAN Native AI}
\label{sec:conflict_native_ai}
The O-RAN Alliance has placed {Native \ac{AI}} at the core of its architecture. It exposes native interfaces (A1, \ac{E2SM}-\ac{EI}, \ac{AIMLFW}) so that data collection, model training and inference can be embedded within the \ac{RAN} if desired. These \ac{AI}/\ac{ML} blocks remain optional and may also reside in an external MLOps platform that interfaces with the RIC via the same open interfaces. Samsung's \ac{AIMLFW} prototype shows how \ac{NWDAF}/\ac{PCF} analytics are converted by the SMO Data-Collection Function and streamed to the training host for \ac{RL} retraining \cite{lee2021onlineRIC}. As highlighted by ~\cite{DryjanskiBlog}, Open \ac{RAN} natively supports \ac{AI}/\ac{ML} through well-defined functional blocks (e.g., \ac{ML} Training Host (MTH) and \ac{ML} Inference Host (MIH)) and interfaces (e.g., E2, O1, A1). In this section, we discuss how the three essential dimensions of conflict management, namely \emph{detection, classification, and mitigation}, can be realized within the broader Open \ac{RAN} \ac{ML} framework.

\subsection{Conflict Detection in the Open RAN ML Workflow}
\subsubsection{Data Collection for Anomaly and Conflict Indicators.}
Open \ac{RAN}’s native \ac{AI} workflow begins by collecting measurements from \ac{O-RU}, \ac{O-DU} and \ac{O-CU} over the O1 interface or via \ac{E2SM}-\ac{KPM}/UEINF, while the SMO’s \emph{Data-Collection Function} (DCF) ingests Core-Network analytics (NWDAF/PCF) and application-function KPIs exposed through NEF. The DCF normalizes every feed- including UE-level counters observed at the gNB-into standard \emph{Enrichment-Information} records before forwarding them to the \ac{AIMLFW}; thus the framework never interfaces with UEs directly, a workflow explicitly outlined in the O-RAN WG 2 technical report on A1 policy-conflict mitigation \cite{ORAN_A1PCM_TR}.

In the context of \emph{conflict detection}, this rich data (which can include \acp{KPI} and contextual \ac{EI}), such as - \ac{RCP}, \ac{P2X}, \ac{P2K}, \ac{K2X}, and \ac{VK}, is vital. Conventional threshold-based conflict detection triggers alarms post-impact, but \ac{AI}-based systems within the Non-RT RIC or \ac{Near-RT-RIC} can proactively detect the \emph{onset} of conflicts using:
\begin{itemize}[leftmargin=*]
  \item \emph{Unsupervised learning models}: Clustering or density-estimation algorithms (e.g., K-means, Isolation Forest) trained on normal network states to flag anomalies indicative of \ac{xApp} conflicts.
  \item \emph{Time-series analysis}: LSTM or Transformer-based architectures, leveraging the broad Open \ac{RAN} data collection, to predict unusual \ac{KPI} patterns before severe performance degradation occurs.
\end{itemize}

\subsubsection{Proactive Alerts and Feedback Loops.}
Once a potential conflict is detected by the MIH, it can generate an alert for the Actor (e.g., \ac{O-DU}, \ac{O-CU}, or \ac{Near-RT-RIC}) to adjust configuration parameters via the E2 or O1 interface. This proactive loop ensures that conflicts are addressed at an early stage, minimizing adverse effects on network performance. 

\subsection{Conflict Classification Leveraging AI at Non-RT and Near-RT-RIC}
The classification can be done for both the binary, non-binary cases. Binary classification between conflict and no-conflict or multiclass classification among three defined conflict types and no conflict can be considered. In our proposed \ac{CDC}, the Stage-1 (non-AI, detection) raises a binary alarm to detect whether there is a conflict or not, and Stage-2 (AI-based, classification) refines that alarm into \textit{direct}/\textit{indirect}/\textit{implicit}. The same hook allows future labels--e.g., policy-inconsistency, resource-contention, life-cycle overlap to be introduced with no change to the real-time detector; only the classifier and the \ac{CMC}’s mitigation-lookup table need re-training.

\subsubsection{Scalability of Classification.}  
As Open \ac{RAN} scales with dozens--and soon even \textit{hundreds} of \acp{xApp},\footnote{Operator-grade RIC clusters are already dimensioned for “\textit{potentially hundreds of xApps}” in dense deployments~\cite{cormoran25}.   Ericsson’s public rApp Directory markets itself as a one-stop shop for \textit{hundreds of rApps}~\cite{ericssonRappDir24}.   Industry analysts at SNS Telecom count \textit{more than 50 independent vendors} actively building commercial xApps/rApps, with rapid year-on-year growth~\cite{snstelecom24}.}  rule-based classification becomes impractical. \ac{AI}-driven methods provide a scalable alternative, leveraging models like GNNs to capture \ac{xApp} dependencies and LSTMs to analyze temporal patterns.  The deployment of classification models depends on control loop constraints: 

\begin{itemize}[leftmargin=*]
  \item \emph{Non-RT RIC}: Suitable for high-latency, offline learning with extensive computational resources.
  \item \emph{Near-RT RIC}: Executes conflict-classification models inside the 10ms to 1s control-loop window mandated by O-RAN Alliance for near-real-time functions and returns a new policy to the DU via the E2 interface typically within \(\approx\!100\;\text{ms}\) in our study.\footnote{O-RAN Alliance specifies 10ms as the lower bound for near-RT control loops and reserves sub-10ms “hard real-time” operation for \acp{dApp} embedded in the RU/DU. Our classifier therefore targets the 10ms to 1s near-RT budget, not the \ac{dApp} micro-second requirement.}
\end{itemize}

\subsubsection{Online Learning and Continuous Adaptation.}
\label{subsec:online_learning}
Open \ac{RAN}’s \ac{ML} framework foresees \emph{online learning}, allowing classification models to update continuously as new patterns of conflicts emerge. This adaptation is critical when \acp{xApp} and traffic profiles evolve, ensuring classification remains accurate without extensive manual rule maintenance. In our architecture, continuous adaptation is realized via periodic or drift-triggered retraining in the Non-RT RIC. Newly collected KPI/EI samples are buffered in the Training Cache; once a retraining threshold (e.g., time-based or volume-based) is met, the conflict-classification model is re-trained, packaged as a new inference bundle, and registered in the AI Model Catalog. After passing governance checks, this updated model version is re-deployed to the Near-RT RIC, replacing the previous one without changing the real-time control logic.

\subsection{Conflict Mitigation via AI-Assisted Decision Making}

Traditional game–theoretic solutions (e.g., maximising the Nash Social Welfare, NSW) remain tractable only while the number of competing vendors \(V\) and the set of simultaneously conflicting \acp{xApp} \(X\) are smaller. \footnote{Nguyen \emph{et al.} prove that exact NSW maximisation is NP-hard and that the solver’s variable count grows linearly with the number of agents, i.e.\ \(V\) \cite{nguyen2014complexity}. Wadud \emph{et al.} evaluate a NSW-based Conflict Mitigation Controller on only two xApps and report solver run times well below the 1s near-RT budget, but note that complexity “explodes quickly” with more actors \cite{wadud2023conflict}. Brach del Prever \emph{et al.} show that even sandbox profiling of \emph{ten} xApps requires offline analysis and cannot run inside the near-RT loop \cite{del2024pacifista}. } 
When \(V\) grows to tens and \(X\) to hundreds, the decision space explodes quadratically and solver run times breach the 10ms to 1s near-RT budget.  \ac{AI}-based methods sidestep this by learning an allocation policy offline and issuing a resolution in \(\mathcal{O}(V+X)\) time at run-time.

\begin{itemize}[leftmargin=*]
  \item \emph{Reinforcement Learning (RL)}: \ac{RL} agents \cite{10827667} can observe real-time \ac{KPI} feedback and proactively coordinate resource sharing, ensuring all \acp{xApp} meet respective \ac{QoS} and SLA requirements. An early \ac{PPO}-based \ac{xApp} that adapts \ac{DU} parameters and boosts cell throughput by 19\% after on-line fine-tuning is presented in \cite{lee2021onlineRIC}; we follow the same personalize-then-deploy philosophy for conflict mitigation.
  \item \emph{Multi-Agent \ac{RL} (MARL)}: Treats each \ac{xApp} as an agent, allowing them to negotiate resource allocation autonomously while ensuring system-wide optimization.
  \item \emph{Transformer-based \cite{10615438} or attention-based strategies}: Model global context across many \acp{xApp} simultaneously, identifying resource conflicts and proposing reconfigurations that preserve overall \ac{RAN} performance.
\end{itemize}

\subsubsection{Deployment Scenarios for Mitigation Models.}
Similar to detection and classification, conflict-mitigation logic can can be deployed:
\begin{itemize}[leftmargin=*]
  \item \emph{On the Non-RT RIC} (rApps), where it computes long-term or
        bulk policies; these are enforced either  
        (i) by sending an \texttt{A1-P/ML} message to a companion policy-enforcer xApp in the Near-RT RIC, which translates the policy into \texttt{E2-SET/INSERT} commands, or  (ii) by pushing updated YANG objects through the SMO/O1 interface when sub-second reaction is not required.
  \item \emph{On the Near-RT RIC} (xApps), which can execute the mitigation model itself and issue \texttt{E2} commands directly, meeting the 10ms to 1s near-RT control-loop budget.
\end{itemize}
This placement aligns with the Open RAN ML deployment options (e.g., Option 2 in \cite{ORANML}), ensuring each step from detection to mitigation is executed in the most appropriate control loop. We note that the architectural question of placing mitigation logic in the Non-RT versus Near-RT RIC has already been examined extensively in prior O-RAN deployment studies \cite{10287529,10329927}, and our discussion here serves only to contextualize where the proposed CMS algorithms would execute rather than to re-evaluate deployment architectures.

\subsection{Native AI, Continuous Feedback, and End-to-End Conflict Management}

\textit{Closing the Loop:}
Once a mitigation action is taken, the \emph{Subject} (e.g., \ac{O-CU}, \ac{O-DU}, or \ac{O-RU}) provides updated measurements back to the \ac{ML} training or inference hosts, completing the feedback loop. If the mitigation approach fails to resolve the conflict or inadvertently spawns new ones, the \ac{AI} models can adapt based on revised \ac{KPI} feedback and re-train or refine policies accordingly. Model refinement follows the continuous-update workflow outlined in Sec.~\ref{subsec:online_learning}.

\textit{End-to-End View:}
By embedding conflict detection, classification, and mitigation within the broader Open \ac{RAN} \ac{ML} framework, operators achieve a fully automated, \emph{end-to-end} solution. Conflicts are spotted before they escalate, classified in near-real time, and resolved using data-driven policies that balance \ac{xApp} \ac{QoS} demands with overall network performance.

\section{Synthetic Dataset Generation with \texttt{GenC}}
\label{sec:genc}
To train and benchmark the AI classifiers we need a large, \emph{labeled} stream of xApp interactions. Live testbeds rarely exhibit all conflict modes, so we build the dataset synthetically with our \texttt{GenC} generator.\footnote{GenC extends the stochastic model proposed in \cite{wadud2023conflict,wadud2024qacm}; meta-datasets are available at \url{https://github.com/dewanwadud1/GenC}.} Below we give a high-level view of the synthetic data generation process.

\subsection{Entity synthesis and relationship maps}
\begin{enumerate}[leftmargin=*]
\item \textit{Enumerate ICPs and KPIs.}  
      For $M$ xApps we create  
      $\displaystyle P = 2M+\!\left\lfloor M/2\right\rfloor$ ICPs and
      $K = P/2$ KPIs, matching the “$\sim2000$ parameters per gNB” scale
      reported in \cite{ORANSC2025}.  Each symbol is named
      $p_i$ or $k_j$ to keep the feature vectors compact.
\item \textit{Unique seeding.}  
      Every xApp $x_m$ is first given one exclusive ICP and one exclusive
      KPI; this guarantees \emph{direct} influences exist.
\item \textit{Controlled sharing ($p=0.3$).}  
      Remaining parameters and KPIs are re-drawn and assigned with
      probability $p$ to \emph{two} distinct xApps, creating realistic
      \emph{shared-control} surfaces without flooding the graph.
      The result is the mapping tables P2X and K2X.
\item \textit{KPI-parameter groups.}  
      For every KPI we union the ICPs of all xApps that manage it,
      obtaining the set
      $\text{P}_{k_j}\!\subseteq\!\{p_1,\dots,p_P\}$ and thus P2K.
\item \textit{Indirect-conflict injection.}  
      To emulate latent couplings, one parameter from a \emph{different}
      xApp is randomly added to each
      $\text{P}_{k_j}$.  This produces
      indirect-conflict tuples without breaking exclusivity.
\item \textit{Implicit parameters.}  
      At least one ICP is kept unassigned; its random tweaks later
      degrade KPIs via the Gaussian channel, yielding implicit
      conflicts.
\end{enumerate}

These six steps give the static overlays P2X, P2K, K2X, the
\emph{shared-parameter graph} $G_{\text{XP}}$ and the
\emph{KPI-coupling graph} $G_{\text{KP}}$ (Algorithms
\texttt{generate\_xp\_graph} and \texttt{generate\_kp\_graph} \cite{wadud2024xapp}).  They
are stored once and reused during simulation.

\subsection{Time-series simulation loop}
For each second $t\in[1,T]$ we:

\textit{(i) pick an ICP} from three buckets (i.e., shared, unassigned, or indirect) with specific probability on each parameter to mirror field-log statistics \cite{i14y};  

\textit{(ii) update its value} by a random drift
$\Delta p\!\sim\!\mathcal{U}(-70,70)$;  

\textit{(iii) recompute affected KPIs} via a Gaussian response
\[
k_j(t)\;=\;\exp\!\left[-\frac{(p_i+\xi)^2}{2\sigma^2}\right],
\qquad\xi\!\sim\!\mathcal{U}(-20,20),
\]
which captures the local, non-linear sensitivity of a KPI to its parameters that mimics their real-world properties \ac{RAN} \cite{banerjee2021toward};

\textit{(iv) compare} $k_j(t)$ with its SLA threshold
$\tau_j\!\sim\!\mathcal{U}(0.6,0.8)$ and store only the \emph{new}
violations, mimicking the near-RT alarm stream; we use different SLA bounds for different conflicts intensities.

\textit{(v) label the row} with the rule engine of
Alg.\,\ref{alg:rule_based_mitigation}, producing the ground-truth class
\textit{No/Direct/Indirect/Implicit Conflict}.

All features, including the RCP string, parameter/KPI snapshots, mapping columns
and conflict labels are appended in a single CSV line. 

\subsection{Realism and class balance}
GenC reproduces three key properties observed in operator logs
\cite{i14y,ORAN_A1PCM_TR}:

\begin{itemize}[leftmargin=*]
\item \textit{Rare events.}  
      Only $<10\,\%$ of rows carry a conflict label, matching the
      “sub-1\% KPI-breach” rate in live RANs \cite{onlineLearningRCA}.
\item \textit{Skewed taxonomy.}  
      Direct:Indirect: Implicit ratios are
      roughly $3:5:2$. This is because parameter changes made by one 
      xApp frequently affects KPIs governed by other xApps through shared 
      dependencies, whereas direct conflicts arise only when both the parameter 
      and the violated KPI belong to the same xApp, and implicit conflicts are 
      triggered only through secondary effects. A recent study tests models on a synthetic dataset where only 10-40\% of the instances involve actual conflicts, and the rest of the time (90-60\%), everything runs smoothly \cite{onlineLearningRCA}.
\item \textit{Heterogeneous influence graph.}  
      Only a handful of ICPs are shared by many xApps, while the vast
      majority are unique to one or two controllers, resulting in a
      \emph{heavy-tailed} degree distribution for $G_{\text{XP}}$.
      This “few-parameters-highly-shared” pattern is qualitatively consistent with the parameter-sharing observations reported in the PACIFISTA open-RAN test cluster \cite{del2024pacifista}.
\end{itemize}
Consequently, the datasets drive the classifier under the same class-
imbalance and structural sparsity it will face in production, while
remaining fully reproducible and parameterizations for ablation studies.

\section{Experiment and Results}
\label{sec:simRes}
To verify the potential of \ac{ML} in providing scalable conflict classification, we generated datasets for 5, 10, 20, 30, and 50 \acp{xApp}, each spanning one million time steps, with rule-based annotation active, used as ground truth in training the AI/ML models during simulation. The synthetic conflict generation framework is discussed in Section~\ref{sec:genc}. We ensured that each conflict type occurred sufficiently to facilitate \ac{AI} model training. Due to the randomized parameter changes, the generated dataset exhibited class imbalance, closely resembling real-world network conditions. We trained Bi-LSTM and GNN with GraphSAGE on these datasets under imbalanced conditions and compared their performance with SMOTE-GNN (Synthetic Minority Over-sampling Technique for Graph Neural Networks). The architectures of these models are discussed in Section~\ref{sec:aiClassification}. All experiments were executed on a standard Apple MacBook Pro (M1~Pro, 16GB unified memory, integrated 14-core GPU), and the latency values reported correspond to per-sample inference time. Since inference requires only lightweight tensor operations, these runtime remain well within the 10ms-1s Near-RT RIC control-loop budget and are therefore representative of what can be achieved on typical Near-RT RIC servers.


\begin{figure}[!ht]
 \centering
\includegraphics[scale=0.45]{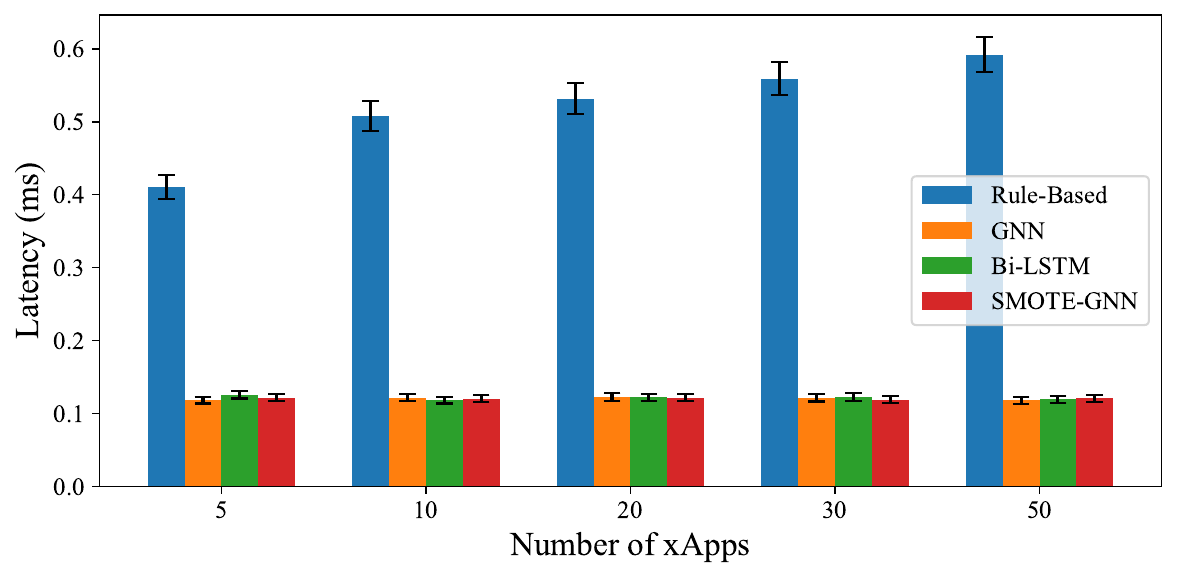}
	\vspace{-0.2in}
	\caption{Average computation time for classifying a conflict}
	\label{fig:compTime}
 \vspace{-0.1in}
\end{figure}

The computational efficiency of different conflict classification methods is illustrated in Figure~\ref{fig:compTime}. The error bars in Fig.~\ref{fig:compTime} are very small, which means the measured computation times are stable across repeated runs and the differences between rule-based and AI methods are not due to random fluctuations. The rule-based method exhibits the highest computational cost across all \ac{xApp} configurations, with classification times remaining relatively high. This is expected as the rule-based approach exhaustively checks predefined conditions, leading to increased complexity as the number of \acp{xApp} grows. In contrast, GNN, Bi-LSTM, and SMOTE-GNN demonstrate significantly lower computational times, with all three achieving a near-constant classification time as the number of \acp{xApp} increases. Notably, Bi-LSTM and SMOTE-GNN perform similarly, but GNN maintains a slightly lower computational cost. This highlights the advantage of graph-based learning for structural conflict detection, where feature aggregation minimizes redundant computations. The results indicate that \ac{AI}-driven methods, particularly GNN-based approaches, offer substantial efficiency gains over traditional rule-based classification.

\begin{table*}[t]
\centering
\caption{Accuracy (\%) of conflict-detection methods across xApp counts and conflict intensities (mean ~$\pm$~ s.e. of 10 runs).}
\label{tab:acc-all-scenarios}
\resizebox{\textwidth}{!}{%

\begin{tabular}{lccccc}
\toprule
 & \textbf{5 xApps} & \textbf{10 xApps} & \textbf{20 xApps} & \textbf{30 xApps} & \textbf{50 xApps} \\
\midrule
\textbf{Rule-Based} & 100 & 100 & 100 & 100 & 100 \\  
\midrule
\textbf{GNN} \\
\quad Low conflict    & 99.87~$\pm$~0.05 & 99.75~$\pm$~0.06 & 99.81~$\pm$~0.05 & 99.85~$\pm$~0.05 & 99.89~$\pm$~0.04 \\
\quad Medium conflict & 97.66~$\pm$~0.14 & 97.04~$\pm$~0.15 & 97.06~$\pm$~0.15 & 97.54~$\pm$~0.14 & 97.67~$\pm$~0.13 \\
\quad High conflict   & 96.28~$\pm$~0.18 & 95.71~$\pm$~0.19 & 95.75~$\pm$~0.19 & 96.10~$\pm$~0.18 & 96.34~$\pm$~0.17 \\
\midrule
\textbf{Bi-LSTM} \\
\quad Low conflict    & 99.90~$\pm$~0.04 & 99.88~$\pm$~0.04 & 99.84~$\pm$~0.05 & 99.86~$\pm$~0.04 & 99.91~$\pm$~0.03 \\
\quad Medium conflict & 98.50~$\pm$~0.11 & 98.37~$\pm$~0.11 & 98.54~$\pm$~0.11 & 98.64~$\pm$~0.10 & 98.68~$\pm$~0.10 \\
\quad High conflict   & 97.21~$\pm$~0.15 & 97.51~$\pm$~0.14 & 97.24~$\pm$~0.15 & 97.58~$\pm$~0.14 & 97.51~$\pm$~0.14 \\
\midrule
\textbf{GNN-SMOTE} \\
\quad Low conflict    & 100.00~$\pm$~0.00 & 100.00~$\pm$~0.00 & 100.00~$\pm$~0.00 & 100.00~$\pm$~0.00 & 100.00~$\pm$~0.00 \\
\quad Medium conflict & 99.99~$\pm$~0.01 & 99.98~$\pm$~0.01 & 99.99~$\pm$~0.01 & 99.99~$\pm$~0.01 & 99.98~$\pm$~0.01 \\
\quad High conflict   & 99.97~$\pm$~0.02 & 99.96~$\pm$~0.02 & 99.97~$\pm$~0.02 & 99.97~$\pm$~0.02 & 99.96~$\pm$~0.02 \\
\bottomrule
\end{tabular}}
\end{table*}

\begin{table*}[t]
\centering
\caption{Macro-averaged F1 (\%) under the same conditions as Table~\ref{tab:acc-all-scenarios} (mean~$\pm$~s.e. of 10 runs).}
\label{tab:f1-all-scenarios}
\resizebox{\textwidth}{!}{%
\begin{tabular}{lccccc}
\toprule
 & \textbf{5 xApps} & \textbf{10 xApps} & \textbf{20 xApps} & \textbf{30 xApps} & \textbf{50 xApps} \\
\midrule
\textbf{GNN} \\
\quad Low conflict    & 93.23~$\pm$~0.05 & 93.27~$\pm$~0.06 & 96.54~$\pm$~0.05 & 97.64~$\pm$~0.05 & 98.49~$\pm$~0.04 \\
\quad Medium conflict & 71.71~$\pm$~0.14 & 66.61~$\pm$~0.15 & 70.08~$\pm$~0.15 & 77.78~$\pm$~0.14 & 80.78~$\pm$~0.13 \\
\quad High conflict   & 71.81~$\pm$~0.18 & 68.96~$\pm$~0.19 & 71.26~$\pm$~0.19 & 75.52~$\pm$~0.18 & 78.49~$\pm$~0.17 \\
\midrule
\textbf{Bi-LSTM} \\
\quad Low conflict    & 94.50~$\pm$~0.04 & 96.82~$\pm$~0.04 & 97.15~$\pm$~0.05 & 97.86~$\pm$~0.04 & 98.78~$\pm$~0.03 \\
\quad Medium conflict & 82.88~$\pm$~0.11 & 83.07~$\pm$~0.11 & 86.38~$\pm$~0.11 & 88.38~$\pm$~0.10 & 89.60~$\pm$~0.10 \\
\quad High conflict   & 79.69~$\pm$~0.15 & 83.26~$\pm$~0.14 & 82.36~$\pm$~0.15 & 85.59~$\pm$~0.14 & 85.93~$\pm$~0.14 \\
\midrule
\textbf{GNN-SMOTE} \\
\quad Low conflict    & 100.00~$\pm$~0.00 & 100.00~$\pm$~0.00 & 100.00~$\pm$~0.00 & 100.00~$\pm$~0.00 & 100.00~$\pm$~0.00 \\
\quad Medium conflict & 99.89~$\pm$~0.01 & 99.79~$\pm$~0.01 & 99.91~$\pm$~0.01 & 99.92~$\pm$~0.01 & 99.83~$\pm$~0.01 \\
\quad High conflict   & 99.76~$\pm$~0.02 & 99.74~$\pm$~0.02 & 99.78~$\pm$~0.02 & 99.80~$\pm$~0.02 & 99.78~$\pm$~0.02 \\
\bottomrule
\end{tabular}}
\end{table*}

Table~\ref{tab:acc-all-scenarios} and Table~\ref{tab:f1-all-scenarios} summarise how each method behaves across different numbers of xApps and conflict intensities (low, medium, high). As expected, the rule-based method always achieves 100\% accuracy because it directly implements the ground-truth heuristics. The AI-based models also reach very high accuracy in all settings: GNN is slightly less accurate than Bi-LSTM and GNN-SMOTE, especially at higher conflict intensities, while Bi-LSTM generally tracks the rule-based performance closely. GNN-SMOTE consistently achieves almost perfect accuracy, showing that handling class imbalance makes a real difference when conflicts become more frequent. The ``$\pm$'' values in both tables are the standard error over 10 independent runs and remain very small, which confirms that the results are repeatable and not driven by outliers.

Macro-averaged F1 in Table~\ref{tab:f1-all-scenarios} is especially important because it treats all classes equally, including rare conflict types. Here, we see that standard GNN and Bi-LSTM lose more performance than their accuracy alone would suggest, particularly in medium and high conflict settings where the minority classes are harder to learn. In contrast, GNN-SMOTE keeps macro-F1 scores close to 100\% across all xApp counts and conflict intensities, which means it correctly recognizes not only the dominant ``no-conflict'' cases but also the less frequent conflict types. In simple terms, accuracy tells us that the AI models almost never make mistakes overall, while macro-F1 shows that GNN-SMOTE is the most reliable model when we also care about the rare but important conflicts. 


\begin{table*}[t]
    \centering
    \caption{Impact of GenC parameters and conflict intensity on classifier performance (accuracy~$\pm$~s.d.).}
    \label{tab:genc_parameters_vs_performance}
    \begin{tabular}{|l|l|p{6.3cm}|ccc|}
        \toprule
        & & & \multicolumn{3}{c}{Average accuracy [\%]} \\
        \cmidrule(lr){4-6}
        Conflict intensity 
        & Conflict ratio range [\%] 
        & GenC configuration (structural and dynamic factors) 
        & GNN 
        & Bi-LSTM 
        & GNN-SMOTE \\
        \midrule
        Low 
        & 1--4 
        & \textbf{ICP probabilities:} 
          \newline shared ICPs $= 0.30$, indirect ICPs $= 0.50$, unassigned ICPs $= 0.20$.
          \newline \textbf{Dynamic GenC factors for Low:}
          \newline -- Parameter-update frequency $= 0.05$
          \newline -- Relaxed SLA band, $= 0.30$ 
          \newline -- Per-update breach probability $= 0.03$
        & 99.83~$\pm$~0.05 
        & 99.88~$\pm$~0.04 
        & 100.00~$\pm$~0.00 \\
        \addlinespace
        Medium 
        & 5--7 
        & \textbf{ICP probabilities:} 
          \newline shared ICPs $= 0.30$, indirect ICPs $= 0.50$, unassigned ICPs $= 0.20$.
          \newline \textbf{Dynamic GenC factors for Medium:}
          \newline -- Parameter-update frequency $= 0.10$
          \newline -- Relaxed SLA band, $= 0.20$ 
          \newline -- Per-update breach probability $= 0.06$
        & 97.39~$\pm$~0.14 
        & 98.55~$\pm$~0.11 
        & 99.99~$\pm$~0.01 \\
        \addlinespace
        High 
        & 8--10 
        & \textbf{ICP probabilities:} 
          \newline shared ICPs $= 0.30$, indirect ICPs $= 0.50$, unassigned ICPs $= 0.20$.
          \newline \textbf{Dynamic GenC factors for High:}
          \newline -- Parameter-update frequency $= 0.15$
          \newline -- Relaxed SLA band, $= 0.15$ 
          \newline -- Per-update breach probability $= 0.10$
        & 96.04~$\pm$~0.18 
        & 97.41~$\pm$~0.14 
        & 99.97~$\pm$~0.02 \\
        \bottomrule
    \end{tabular}
\end{table*}

Table~\ref{tab:genc_parameters_vs_performance} shows how classifier performance changes as we vary the GenC parameters that control the difficulty of the generated conflict patterns. Low conflict intensity corresponds to infrequent KPI breaches and relaxed SLA bounds, while high intensity reflects more frequent parameter updates, tighter SLA margins, and a greater likelihood of generating conflicting behaviour. As the conflict intensity increases from low to medium to high, all models experience a gradual drop in accuracy, which is expected because the data become more challenging and increasingly imbalanced. Among the three methods, GNN-SMOTE remains the most stable across all conditions, followed by Bi-LSTM and then standard GNN. This demonstrates that GenC's structural and dynamic parameters directly influence task difficulty and that addressing class imbalance is essential for maintaining high classifier reliability under more demanding conflict scenarios.

\section{Evaluation in ns3-oran}
\label{sec:ns3-oran}
In this section, we evaluate our proposed conflict classification method in the ns3-oran \cite{ns3oran_github} simulator. The simulation is performed on a topology based on an open-source cell database, OpenCelliD \cite{opencellid}. 

\begin{figure}[!ht]
 \centering
\includegraphics[scale=0.6]{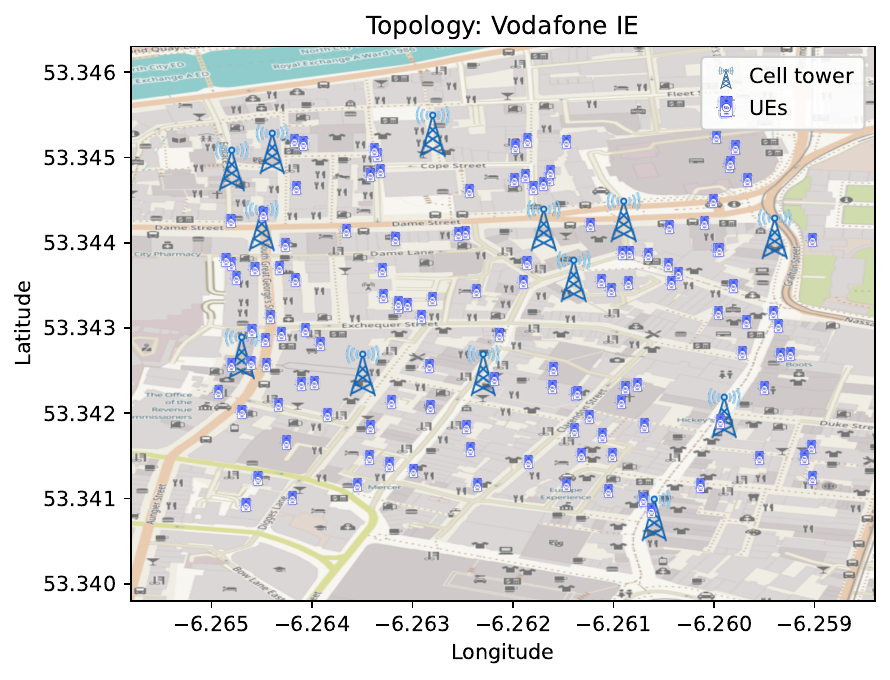}
	\vspace{-0.1in}
	\caption{Cell Topology of Dublin City Center in a $0.4 \times 0.5$ sq. km. area within latitude and longitude $[53.343, -6.262]$ for Vodafone IE.}
	\label{fig:topology}
 \vspace{-0.1in}
\end{figure}

\par Using OpenCelliD (MCC 272, MNC 1) as a source, we filtered LTE records to a Dublin bounding box (53.320--53.356$^\circ$ N, $-6.305^\circ$ to $-6.187^\circ$). Afterwards, we cropped a 0.4 $\times$ 0.5 km sub-window centered at 53.343$^\circ$ N, $-6.262^\circ$ E. Finally, we estimated the coverage radius from each cell's reported range. The same projected positions were exported as an NS-3 \textit{ListPositionAllocator} snippet to use in an ns3-oran simulation setup. The Plot in Fig.~\ref{fig:topology} includes corner latitude/longitude for traceability.


\par There are 13 LTE cells located in the selected geographical area, and we considered 117 UEs (9 per cell) for initial positioning.

\subsection{ns3-oran Simulation Setup and KPI Data Collection}

Our ns3-oran simulation was configured with the OpenCelliD-derived Dublin topology comprising 13 eNodeBs and 117 UEs to generate realistic conflict scenarios between ES and MRO xApps. The simulation focused on collecting comprehensive network \ac{KPI} data while systematically varying \ac{ICP} values, particularly transmission power (TXP), to induce authentic conflicts between energy optimization and mobility management objectives.

The data collection process spanned a total simulation time of 540 seconds. During this period, the simulation monitored six critical \acp{KPI}: Energy Efficiency, Power Consumption, Throughput, Handover Rate, Call Drop Rate, and Handover Failure Rate. Simultaneously, six key \acp{ICP} were tracked: Transmission Power (TXP), Time to Trigger (TTT), Cell Individual Offset (CIO), Neighbor List (NL), Hysteresis (HYS), and Radio Electrical Tilt (RET).


Each timestamped entry in the collected dataset represents a complete snapshot of the network state, including parameter values, KPI measurements, Service Level Agreement (SLA) thresholds, and the resulting conflict classification. The ES xApp controlled energy-related parameters to minimize power consumption, while the MRO xApp optimized mobility parameters to reduce handover failures and call drops. This created realistic conflict scenarios where energy-saving actions (e.g., reducing TXP) conflicted with mobility optimization requirements (e.g., maintaining adequate signal strength for seamless handovers).

The resulting dataset contains detailed conflict annotations generated using our rule-based classification algorithm (Algorithm~\ref{alg:rule_based_mitigation}). Each row includes the conflict type classification (Direct or No Conflict), the instructing xApp, the violated KPIs, and comprehensive relationship mappings (P2X, P2K, K2X), which are essential for both rule-based and AI-based conflict detection methods. This dataset serves as the foundation for comparing traditional rule-based approaches against our proposed AI-driven classification techniques.

\subsection{ES/MRO Conflict Detection Performance Analysis}

To validate our AI-powered conflict management framework in a realistic Open RAN environment, we implemented and evaluated the proposed methods using the collected ns3-oran simulation data. We specifically focused on the ES and MRO xApp conflict scenario, as these represent one of the most common and critical conflict types in practical Open RAN deployments.

Our evaluation compared four conflict detection and classification approaches: Rule-Based (Algorithm~\ref{alg:rule_based_mitigation}), Bi-LSTM, GNN, and SMOTE-GNN. The collected NS-3 dataset provides authentic conflict scenarios where energy optimization decisions impact mobility performance and vice versa, with TXP serving as the primary shared parameter that causes direct conflicts between the two xApps.

\begin{figure*}[!t]
\centering
\includegraphics[width=0.9\textwidth]{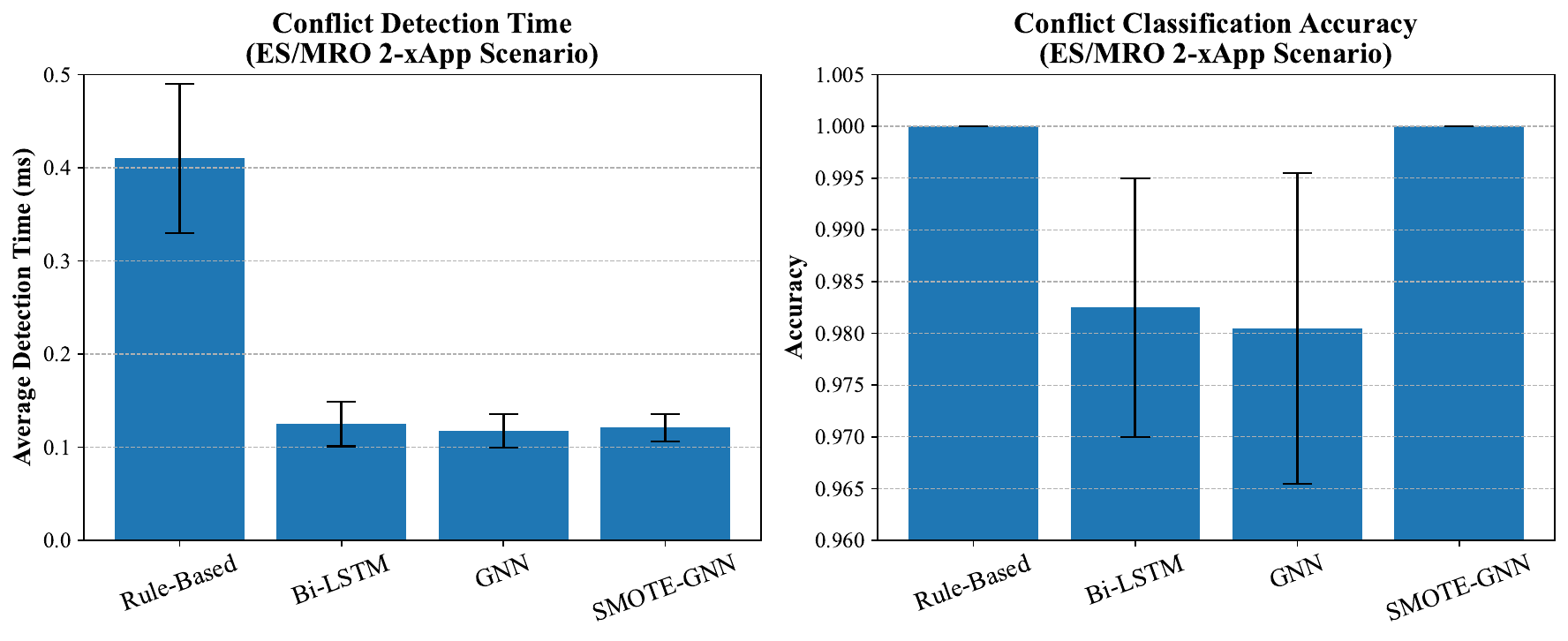}
\caption{Performance comparison of conflict detection methods in ns3-oran simulation: (a) Average detection time comparison showing AI methods achieve significantly faster classification than rule-based approaches, (b) Classification accuracy comparison demonstrating high accuracy across all methods, with SMOTE-GNN achieving perfect classification.}
\label{fig:ns3_performance_comparison}
\end{figure*}

The results, presented in Figure~\ref{fig:ns3_performance_comparison}, demonstrate the significant advantages of AI-based conflict classification over traditional rule-based methods. The error bars shown in Fig.~\ref{fig:ns3_performance_comparison} are very tight for both detection time and accuracy, indicating that these gains are consistent across runs and not the result of noisy measurements. The detection time analysis (Figure~\ref{fig:ns3_performance_comparison}a) reveals that the rule-based method requires 0.410 ms per classification, while AI-based approaches achieve substantially faster performance: GNN (0.118 ms), SMOTE-GNN (0.121 ms), and Bi-LSTM (0.125 ms). This represents an average speedup of $3.2\times$ for AI methods, which is crucial for meeting Near-RT-RIC's 10ms-1s control loop requirements.

Regarding classification accuracy (Figure~\ref{fig:ns3_performance_comparison}b), all methods demonstrate excellent performance. The rule-based approach achieves 100\% accuracy by design, as it follows deterministic classification rules. Among AI methods, SMOTE-GNN achieves perfect 100\% accuracy, demonstrating the effectiveness of synthetic minority oversampling for handling imbalanced conflict datasets. Bi-LSTM also achieves 100\% accuracy, while standard GNN reaches 99.05\% accuracy, indicating minor sensitivity to class imbalance in the ES/MRO scenario.

These ns3-oran simulation results validate our synthetic dataset findings and confirm that AI-based conflict classification can effectively handle real-world ES/MRO scenarios. The consistent performance across both synthetic (Table~\ref{tab:acc-all-scenarios}) and ns3-oran environments demonstrates the robustness and practical applicability of our proposed AI-powered conflict management framework. Particularly, SMOTE-GNN emerges as the optimal solution, combining the fastest detection time among AI methods with perfect classification accuracy, making it ideal for large-scale Open RAN deployments with multiple vendor xApps.

\subsection{Conflict Mitigation Performance}
\label{sec:conflictMitiPerformance}

\begin{figure}[!t]
\centering
\includegraphics[width=0.48\textwidth]{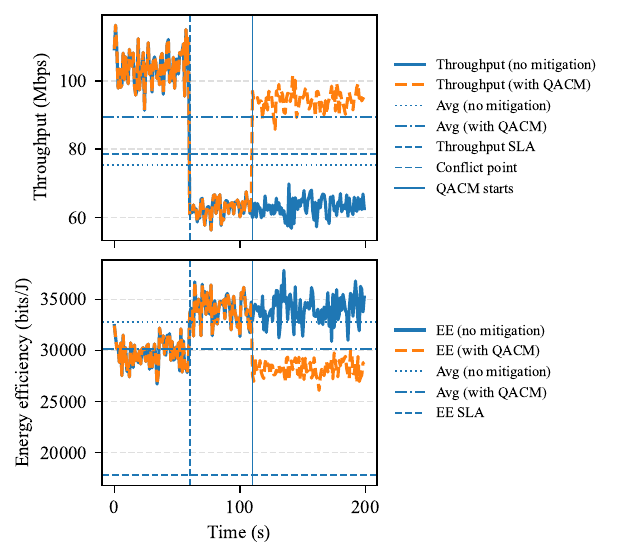}
\caption{Time-series of a direct ES-MRO conflict and its mitigation via QACM \cite{wadud2024qacm} in the ns3-oran simulation. A reduction in transmission power by the ES xApp pushes throughput below its SLA threshold, triggering a direct conflict. Once activated, QACM adjusts the power level to a compromise point, restoring both throughput and energy-efficiency KPIs above their SLA bounds.}
\label{fig:ns3_QACM}
\end{figure}

Fig.~\ref{fig:ns3_QACM} shows an instance of a direct conflict between the ES and MRO xApps and how it is resolved using the QACM method \cite{wadud2024qacm}. The KPIs remain stable at the start of the simulation until the ES xApp reduces the transmission power to save energy in the selected window in Fig.~\ref{fig:ns3_QACM}. This action suddenly lowers the throughput below the SLA threshold, creating a direct conflict with the MRO xApp. Without mitigation, both the throughput and energy-efficiency KPIs remain imbalanced. The throughput remains below its SLA threshold for 10 continuous seconds, which is the trigger time for possible conflict detection and classification component in our simulation. The detection and classification confirms a direct conflict over transmission power and triggers the mitigation component.

When QACM is activated, it reacts to the detected violation and adjusts the transmission power to a safe compromise point by running its optimization. After this adjustment, both KPIs recover and move above their respective SLA thresholds. This demonstrates that QACM can quickly restore KPI balance and maintain service quality during conflicting xApp decisions.

\subsection{Scalability Analysis for Multi-Vendor Open RAN}

The ns3-oran results provide important insights for scaling to multi-vendor Open RAN environments. While our simulation focused on the two-xApp ES/MRO scenario, the performance characteristics align with our synthetic dataset analysis across 5-50 xApps. The consistent sub-millisecond detection times for AI methods suggest that the proposed framework can handle the complexity of modern Open RAN deployments where dozens of xApps from different vendors simultaneously optimize network parameters.

The $3.2\times$ speed improvement over rule-based methods becomes increasingly critical as the number of xApps grows. Based on our complexity analysis in Section~\ref{sec:rule2adaptive}, rule-based classification time scales as $O(n×v×a)$, while AI methods maintain O(1) per-sample complexity through batched tensor operations. This fundamental difference ensures that AI-based classification remains viable even as Open RAN networks scale to hundreds of xApps and thousands of ICPs, as envisioned for 6G deployments. The small error bars and stable macro-F1 scores across all xApp counts and conflict intensities further support that this scalability is not only theoretical but also reflected in robust empirical behaviour.

\section{Conclusion}
\label{sec:conclusion}
Conventional conflict management in Open \ac{RAN} often detects issues too late
and struggles to adapt to the increasing complexity of \ac{xApp} interactions.
In contrast, \ac{AI}-based methods provide a proactive approach by enabling the
network to anticipate conflicts, classify them with high accuracy, and rapidly
mitigate their impact. Our work demonstrates that AI-based conflict
classification achieves $3.2\times$ speed improvement over rule-based methods
while maintaining near perfect accuracy in realistic Open RAN scenarios using ns3-oran. The GenC framework successfully generates training datasets that reflect real-world conflict distributions, enabling robust AI model development.

We acknowledge that the present study evaluates scenarios with up to 50~xApps, which is far smaller than the thousands of ICPs and KPIs envisioned for large-scale O-RAN and future 6G deployments. Scaling to such sizes is currently constrained by the computational cost of GenC dataset generation and the near-quadratic growth of GNN neighborhood sampling as controllable parameters increase. Addressing massive-scale deployments therefore remains an open challenge. Future work will explore distributed and multi-node training, hierarchical and clustered graph representations, and adaptive learning mechanisms to enable scalable CMS operation under the diverse, dynamic, and unpredictable conditions expected in 6G networks.


%



\section*{Acknowledgment}
This research was partially funded by the European Union's Horizon Europe research and innovation program through the Marie Skłodowska-Curie SE grant, under agreement number RE-ROUTE No 101086343.

\ifCLASSOPTIONcaptionsoff
  \newpage
\fi



%
\bibliographystyle{ieeetr}
\bibliography{references}

\appendices
\end{document}

%% file: Acronyms.tex
\begin{acronym}[BEREC]
\acro{10G-EPON}{10 Gigabit EPON}
\acro{3GPP}{3rd Generation Partnership Project}
\acro{5G}{Fifth Generation}
\acro{6G}{Sixth Generation}
\acro{A-CPI}{A Controller Plane Interface}
\acro{A-RoF}{Analog Radio-over-Fiber}
\acro{ABNO}{Application-Based Network Operations}
\acro{ADC}{Analog-to-Digital Converter}
\acro{AE}{Allocative Efficiency}
\acro{amc}{adaptive modulation and coding}
\acro{AON}{Active Optical Network}
\acro{AI}{Artificial Intelligence}
\acro{ML}{Machine Learning}
\acro{API}{Application Programming Interface}
\acro{AWG}{Arrayed Waveguide Grating}
\acro{B2B}{Business to Business}
\acro{B2C}{Business to Consumer}
\acro{B-RAS}{Broadband Remote Access Servers}
\acro{BaaS}{Blockchain as a Service}
\acro{BB}{Budget Balance}
\acro{BBF}{BroadBand Forum}
\acro{BBU}{Base Band Unit}
\acro{BER}{Bit Error Rate}
\acro{BEREC}{Body of European Regulators for Electronic Communications}
\acro{BF}{beamforming}
\acro{BFT}{Byzentine fault tolerant}
\acro{BGP}{Border Gateway Protocol}
\acro{BMap}{Bandwidth Map}
\acro{BPM}{Business Process Management}
\acro{BS}{Base Station}
\acro{BSC}{Base Station Controller}
\acro{BtB}{Back to Back}
\acro{BW}{Bandwidth}
\acro{C2C}{Consumer to Consumer}
\acro{C-RAN}{Cloud Radio Access Network}
\acro{C-RoFN}{Cloud-based Radio over optical Fiber Network}
\acro{CA}{Certificate Authority}
\acro{CapEx}{Capital Expenditure}
\acro{CDMA}{code division multiple access}
\acro{CDR}{Call Drop Rate}
\acro{CFO}{carrier frequency offset}
\acro{CIR}{Committed Information Rate}
\acro{CO}{Central Office}
\acro{CoMP}{Coordinated Multipoint}
\acro{COP}{Control Orchestration Protocol}
\acro{CORD}{Central Office Rearchitected as a Data Centre}
\acro{CPE}{Customer Premises Equipment}
\acro{CPRI}{Common Public Radio Interface}
\acro{CPU}{Central Processing Unit}
\acro{CMC}{Conflict Mitigation Controller}
\acro{CMS}{Conflict Management System}
\acro{CQI}{channel quality indicator}
\acro{CR}{cognitive radio}
\acro{CS}{Central Station}
\acro{CSI}{channel state information}
\acro{D2D}{Device to Device}
\acro{D-CPI}{D Controller Plane Interface}
\acro{D-RoF}{Digital Radio-over-Fiber}
\acro{DAC}{Digital-to-Analog Converter}
\acro{DAS}{distributed antenna system}
\acro{DAS}{Distributed Antenna Systems}
\acro{DBA}{Dynamic Bandwidth Allocation}
\acro{DBRu}{Dynamic Bandwidth Report upstream}
\acro{DC}{Direct Current}
\acro{DL}{Downlink}
\acro{DLT}{Distributed Ledger Technology}
\acro{DMT}{Discrete Multitone}
\acro{DNS}{Domain Name Server}
\acro{DPDK}{Data Plane Development Kit}
\acro{DSB}{Double Side Band}
\acro{DSL}{Digital Subscriber Line}
\acro{DSLAM}{Digital Subscriber Line Access Multiplexer}
\acro{DSP}{Digital Signal Processing}
\acro{DSS}{Distributed Synchronization Service}
\acro{DU}{Distributed Unit}
\acro{DUE}{D2D user equipment}
\acro{DWDM}{Dense Wave Division Multiplexing}
\acro{E-CORD}{Enterprise CORD}
\acro{EAL}{Environment Abstraction Layer}
\acro{EDF}{Erbium-Doped Fiber}
\acro{EFM}{Ethernet in the First Mile}
\acro{EMBS}{Elastic Mobile Broadband Service}
\acro{EON}{Elastic Optical Network}
\acro{EPC}{Evolved Packet Core}
\acro{EPON}{Ethernet Passive Optical Network}
\acro{eMBB}{enhanced Mobile Broadband}
\acro{ETSI}{European Telecommunications Standards Institute}
\acro{EVM}{Error Vector Magnitude}
\acro{FANS}{Fixed Access Network Sharing}
\acro{FASA}{Flexible Access System Architecture}
\acro{FBMC}{filter bank multi-carrier}
\acro{FBMC}{Filterbank Multicarrier}
\acro{FCC}{Federal Communications Commission}
\acro{FDD}{Frequency Division Duplex}
\acro{FDMA}{frequency-division multiple access}
\acro{FEC}{Forward Error Correction}
\acro{FFR}{Fractional Frequency Reuse}
\acro{FFT}{Fast Fourier Transform}
\acro{FPGA}{Field-Programmable Gate Array}
\acro{FSO}{Free-Space-Optics}
\acro{FSR}{Free Spectral Range}
\acro{FTN}{faster-than-nyquist}
\acro{FTTC}{Fiber-to-the-Curb}
\acro{FTTH}{Fiber-to-the-Home}
\acro{FTTx}{Fiber-to-the-x}
\acro{G.Fast}{Fast Access to Subscriber Terminals}
\acro{GEM}{G-PON encapsulation method}
\acro{GFDM}{Generalized Frequency Division Multiplexing}
\acro{GMPLS}{Generalized Multi-Protocol Label Switching}
\acro{GPON}{Gigabit Passive Optical Network}
\acro{GPP}{General Purpose Processor}
\acro{GPRS}{General Packet Radio Service}
\acro{GSM}{Global System for Mobile Communications}
\acro{GTP}{GPRS Tunneling Protocol}
\acro{GUI}{Graphical User Interface}
\acro{HA}{Hardware Accelerator}
\acro{HARQ}{Hybrid-Automatic Repeat Request}
\acro{HD}{half duplex}
\acro{HetNet}{heterogeneous networks}
\acro{IaaS}{Infrastructure as a Service}
\acro{I-CPI}{I Controller Plane Interface}
\acro{I2RS}{Interface 2 Routing System}
\acro{IA}{Interference Alignment}
\acro{IAM}{Identity and Access Management}
\acro{IBFD}{in-band full duplex}
\acro{IC}{Incentive Compatibility}
\acro{ICIC}{inter-cell interference coordination}
\acro{ICT}{information and communications technology}
\acro{IEEE}{Institute of Electrical and Electronics Engineers}
\acro{IETF}{Internet Engineering Task Force}
\acro{IF}{Intermediate Frequency}
\acro{IFFT}{inverse FFT}
\acro{ITS}{Intelligent Transport Systems}
\acro{InP}{Infrastructure Provider}
\acro{IoT}{Internet of Things}
\acro{IP}{Internet Protocol}
\acro{IQ}{In-phase/Quadrature}
\acro{IR}{Individual Rationality}
\acro{IRC}{Interference Rejection Combining}
\acro{ISI}{inter-symbol interference}
\acro{ISP}{Internet Service Provider}
\acro{BFT}{Byzantine-Fault-Tolerant}
\acro{IV}{Intelligent Vehicle}
\acro{ITU}{International Telecommunication Union}
\acro{KPI}{Key Performance Indicator}
\acro{KVM}{Kernel-based Virtual Machine}
\acro{L2}{Layer-2}
\acro{L3}{Layer-3}
\acro{LAN}{Local Area Network}
\acro{LO}{Local Oscillator}
\acro{LOS}{Line Of Sight}
\acro{LR-PON}{Long Reach PON}
\acro{LSP-DB}{Label Switched Path Database}
\acro{LTE-A}{Long Term Evolution Advanced}
\acro{LTE}{Long Term Evolution}
\acro{M-CORD}{Mobile CORD}
\acro{M-MIMO}{massive MIMO}
\acro{M2M}{machine-to-machine}
\acro{MAC}{Medium Access Control}
\acro{ME}{Merging Engine}
\acro{MIMO}{multiple input multiple output}
\acro{MME}{Mobility Management Entity}
\acro{mmWave}{millimeter wave}
\acro{MNO}{mobile network operator}
\acro{MPLS}{Multiprotocol Label Switching}
\acro{MRC}{Maximum Ratio Combining}
\acro{MSC}{Mobile Switching Centre}
\acro{MSP}{Membership Service Providers}
\acro{MSR}{Multi-Stratum Resources}
\acro{MTC}{machine type communication}
\acro{mMTC}{massive Machine Type Communications}
\acro{MVNO}{Mobile Virtual Network Operator}
\acro{MVNP}{mobile virtual network provider}
\acro{MZI}{Mach-Zehnder Interferometer}
\acro{MZM}{Mach-Zehnder Modulator}
\acro{NE}{Network Element}
\acro{NFV}{Network Function Virtualization}
\acro{NFVaaS}{Network Function Virtualization as a Service}
\acro{NG-PON2}{Next-Generation Passive Optical Network 2}
\acro{NGA}{Next Generation Access}
\acro{NLOS}{None Line Of Sight}
\acro{NMS}{Network Management System}
\acro{NRZ}{Non Return-to-Zero}
\acro{NTT}{Nippon Telegraph and Telephone}
\acro{OBPF}{Optical BandPass Filter}
\acro{OBSAI}{Open Base Station Architecture Initiative}
\acro{ODN}{Optical Distribution Network}
\acro{OFC}{Optical Frequency Comb}
\acro{OFDM}{orthogonal frequency-division multiplexing}
\acro{OFDMA}{orthogonal frequency-division multiple access}
\acro{OLO}{Other Licensed Operator}
\acro{OLT}{Optical Line Terminal}
\acro{ONF}{Open Networking Foundation}
\acro{ONU}{Optical Network Unit}
\acro{OOB}{out-of-band}
\acro{OOK}{On-off Keying}
\acro{OpenCord}{Central Office Re-Architected as a Data Center}
\acro{OpEx}{Operating Expenditure}
\acro{OSI}{Open Systems Interconnection}
\acro{OSS}{Operations Support Systems}
\acro{OTT}{Over-the-Top}
\acro{OXM}{OpenFlow Extensible Match}
\acro{P2MP}{Ethernet over point-to-multipoint }
\acro{P2P}{Point-to-Point }
\acro{PAM}{Pulse Amplitude Modulation}
\acro{PAPR}{peak-to-average power rating}
\acro{PAPR}{Peak-to-Average Power Ratio}
\acro{PBMA}{Priority Based Merging Algorithm}
\acro{PCE}{Path Computation Elements}
\acro{PCEP}{Path Computation Element Protocol}
\acro{PCF}{Photonic Crystal Fiber}
\acro{PD}{Photodiode}
\acro{PDCP}{RD Control Protocol}
\acro{PDF}{probability distribution function}
\acro{PGW}{Packet Gateway}
\acro{PHY}{physical Layer}
\acro{PIR}{Peak Information Rate}
\acro{PMD}{Polarization Division Multiplexing}
\acro{PON}{Passive Optical Network}
\acro{POTS}{Plain Old Telephone Service}
\acro{PoET}{Proof of Elapsed Time}
\acro{PoW}{Proof of Work}
\acro{PoS}{Proof of Stake}
\acro{PTP}{Precision Time Protocol}
\acro{PWM}{Pulse Width Modulation}
\acro{QAM}{Quadrature Amplitude Modulation}
\acro{QoE}{Quality of Experience}
\acro{QoS}{Quality of Service}
\acro{QPSK}{Quadrature Phase Shift Keying}
\acro{R-CORD}{Residential CORD}
\acro{RA}{Resource Allocation}
\acro{RAN}{Radio Access Network}
\acro{RAT}{Radio Access Technology}
\acro{RB}{resource block}
\acro{RFIC}{radio frequency integrated circuit}
\acro{RIC}{RAN Intelligent Controller}
\acro{RLC}{Radio Link Control}
\acro{RN}{Remote Node}
\acro{ROADM}{Reconfigurable Optical Add Drop Multiplexer}
\acro{RoF}{Radio-over-Fiber}
\acro{RRC}{Radio Resource Control}
\acro{RRH}{Remote Radio Head}
\acro{RRPH}{Remote Radio and PHY Head}
\acro{RRU}{Remote Radio Unit}
\acro{RSOA}{Reflective Semiconductor Optical Amplifier}
\acro{RU}{Remote Unit}
\acro{Rx}{receiver}
\acro{SC-FDMA}{single carrier frequency division multiple access}
\acro{SCM}{single carrier modulation}
\acro{SD-RAN}{Software Defined Radio Access Network}
\acro{SDAN}{Software Defined Access Network}
\acro{SDMA}{Semi-Distributed Mobility Anchoring}
\acro{SDN}{Software Defined Networking}
\acro{SDR}{Software Defined Radio}
\acro{SFBD}{Single Fiber Bi-Direction}
\acro{SGW}{Serving Gateway}
\acro{SI}{self-interference}
\acro{SIC}{self-interference cancellation}
\acro{SIMO}{Single Input Multiple Output}
\acro{SLA}{Service Level Agreement}
\acro{SMF}{Single Mode Fiber}
\acro{SNMP}{Simple Network Management Protocol}
\acro{SNR}{Signal-to-Noise Ratio}
\acro{SP}{Service Providers}
\acro{Split-PHY}{Split Physical Layer}
\acro{SRI-OV}{Single Root Input/Output Virtualization}
\acro{SU}{secondary user}
\acro{SUT}{System Under Test}
\acro{T-CONT}{Transmission Container}
\acro{TC}{Transmission Convergence}
\acro{TCO}{Total Cost of Ownership}
\acro{TD-LTE}{Time Division LTE}
\acro{TDD}{Time Division Duplex}
\acro{TDM}{Time Division Multiplexing}
\acro{TDMA}{Time Division Multiple Access}
\acro{TED}{Traffic Engineering Database}
\acro{TEID}{Tunnel endpoint identifier}
\acro{TLS}{Transport Layer Security}
\acro{TPS}{Transactions per Second}
\acro{TTI}{transmission time interval}
\acro{TWDM}{Time and Wavelength Division Multiplexing}
\acro{Tx}{transmitter}
\acro{UD-CRAN}{Ultra-Dense Cloud Radio Access Network}
\acro{UDP}{User Datagram Protocol}
\acro{UE}{User Equipment}
\acro{UF-OFDM}{Universally Filtered OFDM}
\acro{UFMC}{Universally Filtered Multicarrier}
\acro{UL}{Uplink}
\acro{UMTS}{Universal Mobile Telecommunications System}
\acro{URLLC}{Ultra-Reliable Low-Latency Communication}
\acro{USRP}{Universal Software Radio Peripheral}
\acro{V2I}{Vehicle to Infrastructure}
\acro{V2V}{Vehicle to Vehicle}
\acro{vBBU}{virtualized BBU}
\acro{vBMap}{Virtual Bandwidth Map}
\acro{vBS}{virtual Base station}
\acro{VCG}{Vickery-Clarke-Groves}
\acro{vCPE}{virtual CPE}
\acro{vCPU}{Virtual Central Processing Unit}
\acro{vDBA}{virtual DBA}
\acro{VDSL2}{Very-high-bit-rate digital subscriber line 2}
\acro{VLAN}{Virtual Local Area Network}
\acro{VM}{Virtual Machine}
\acro{VNF}{Virtual Network Function}
\acro{VNI}{visual networking index}
\acro{VNO}{Virtual Network Operator}
\acro{VNTM}{Virtual Network Topology Manager}
\acro{vOLT}{virtual OLT}
\acro{VPE}{virtual Provider Edge}
\acro{VPN}{Virtual Private Network}
\acro{V2X}{Vehicle to Everything}
\acro{VTN}{Virtual Tenant Network}
\acro{VULA}{Virtual Unblunded Local Access}
\acro{WAN}{Wide Area Network}
\acro{WAP}{Wireless Access Point}
\acro{WCDMA}{wide-band code division multiple access}
\acro{WDM-PON}{Wavelength Division Multiplexing - Passive Optical Network}
\acro{WDM}{Wavelength Division Multiplexing}
\acro{WiMAX}{Worldwide Interoperability for Microwave Access}
\acro{WRPR}{Wired-to-RF Power Ratio}
\acro{XG-PON}{10 Gigabit PON}
\acro{XGEM}{XG-PON encapsulation method}
\acro{XOS}{XaaS Operating System}

\acro{RIC}{RAN Intelligent Controller}
\acro{Near-RT-RIC}{Near Real Time RAN Intelligent Controller}
\acro{Non-RT-RIC}{Non Real Time RAN Intelligent Controller}
\acro{xApp}{Extended Application}
\acro{rApp}{Remote Application}
\acro{dApp}{Distributed Application}
\acro{CMS}{Conflict Management System}
\acro{EG}{Eisenberg-Galle}
\acro{NSWF}{Nash's Social Welfare Function} 
\acro{CMC}{Conflict Mitigation Controller}
\acro{CDC}{Conflict Detection Controller}
\acro{PMon}{Performance Monitoring}
\acro{RCP}{Recently Changed Parameter}
\acro{PGD}{Parameter Group Definition}
\acro{RCPG}{Recently Changed Parameter Group}
\acro{DCKD}{Decision Correlated with KPI Degradation} 
\acro{KDO}{KPI Degradation Occurrences}
\acro{MNO}{Mobile Network Operator}
\acro{SDL}{Shared Data Layer}
\acro{SON}{Self Organising Network}
\acro{SF}{SON Function}
\acro{AM}{Arithmetic Mean}
\acro{PKR}{Parameter and KPI Ranges}
\acro{KPI}{Key Performance Indicator}
\acro{ICP}{Input Control Parameter}
\acro{RF}{Radio Frequency}
\acro{C-RAN}{Cloud Radio Access Network}
\acro{RL}{Reinforcement Learning}
\acro{MARL}{Multi Agent Reinforcement Learning}
\acro{CCO}{Capacity and Coverage Optimization}
\acro{ES}{Energy Saving}
\acro{MRO}{Mobility Robustness Optimization}
\acro{MLB}{Mobility Load Balancing}
\acro{V-RAN}{Virtualized Radio Access Network}
\acro{O-RAN}{Open Radio Access Network}
\acro{D-RAN}{Distributed Radio Access Network}
\acro{5GB}{5G and Beyond}
\acro{O-RU}{Open Radio Unit}
\acro{O-DU}{Open Distribution Unit}
\acro{O-CU}{Open Centralised Unit}
\acro{O-CU-CP}{Open Centralised Unit - Control Plane}
\acro{O-CU-UP}{Open Centralised Unit - User Plane}
\acro{D-SON}{Distributed SON}
\acro{C-SON}{Centralised SON}
\acro{CIO}{Cell Individual Offset}
\acro{CBR}{Call Block Ratio}
\acro{HOR}{Handover Ratio}
\acro{CAN}{Cognitive Autonomous Network}
\acro{CF}{Cognitive Function}
\acro{QACM}{QoS-Aware Conflict Mitigation}
\acro{ANN}{Artificial Neural Network}
\acro{PR}{Polynomial Regression}
\acro{MSE}{Mean Squared Error}
\acro{EVS}{Explained Variance Score}
\acro{CS}{Conflict Supervision}
\acro{TXP}{Tx Power}
\acro{TTT}{Time-To-Trigger}
\acro{CIO}{Cell Individual Offset}
\acro{NL}{Neighbour List}
\acro{RET}{Radio Electrical Tilt}
\acro{HYS}{Handover Hysteresis}
\acro{QACMP}{QACM Priority}
\acro{DRL}{Deep Reinforcement Learning}
\acro{gNB}{Next-generation NodeB}
\acro{RSRP}{Reference Signal Received Power}
\acro{P2K}{parameter-to-KPI}
\acro{P2X}{parameter-to-xApp}
\acro{K2X}{KPI-to-xApp}
\acro{VK}{violating-KPI}
\acro{UP}{Uplink}
\acro{DL}{Downlink}
\acro{AIMLFW}{Open RAN AI/ML framework}
\acro{KPM}{Key Performance Measurement}
\acro{E2SM}{E2 Service Model}
\acro{EI}{Enrichment Information}
\acro{NWDAF}{Network Data Analytics Function}
\acro{PCF}{Policy Control Function}
\acro{PPO}{Proximal Policy Optimization}
\acro{GPU}{Graphics Processing Unit}
\acro{SIMD}{Single Instruction, Multiple Data}
\acro{GNN}{Graph Neural Network}
\acro{TL}{Traffic Load}
\end{acronym}